\documentclass{jfm}
\usepackage{graphicx}
\usepackage{float}
\usepackage{epstopdf, epsfig}
\usepackage{xcolor}

\usepackage{mathtools}

\usepackage{xcolor}
\usepackage{bm}
\usepackage{amsmath}

\usepackage{physics}

\def\e{\boldsymbol{e}}
\def\u{\boldsymbol{u}}

\def\n{\boldsymbol{n}}
\def\t{\boldsymbol{t}}
\def\I{\boldsymbol{I}}
\def\T{\boldsymbol{T}}
\def\U{\boldsymbol{U}}

\def\F{\boldsymbol{F}}

\def \lp {\left(}
\def \rp {\right)}
\def \dtex {\mathrm{d}}
\def \cf {c_f} % temp

\def\dbgamma{\dot{\boldsymbol{\gamma}}}

\def\bsigma{\boldsymbol{\sigma}}

\def\hbsigma{\hat{\boldsymbol{\sigma}}}

\def\Can{\mbox{\it Cu}}   % Carreau number

\title{Self-diffusiophoretic propulsion of a spheroidal particle in a shear-thinning fluid}

\author{Guangpu Zhu\aff{1}\footnote{\label{test}These authors contribute equally to this work.},
	Brandon van Gogh\aff{2,4}\ref{test},
 Lailai Zhu\aff{1},
	On Shun Pak\aff{2,3} \footnote{\label{corresp}Email addresses for correspondance: opak@scu.edu, yiman@pku.edu.cn},
	\and Yi Man\aff{5}\ref{corresp}
	}

\affiliation{	\aff{1}Department of Mechanical Engineering, National University of Singapore, 117575,  Singapore
\aff{2} Department of Mechanical Engineering, Santa Clara University, Santa Clara,  CA 95053, USA
\aff{3} Department of Applied Mathematics, Santa Clara University, Santa Clara,  CA 95053, USA
	\aff{4} Department of Energy Science and Engineering, Stanford University, Stanford, CA 94305, USA
 \aff{5}Department of Mechanics and Engineering Science at College of Engineering, and State Key Laboratory
	for Turbulence and Complex Systems, Peking University, Beijing 100871, PR China }

\begin{document}
\maketitle

\begin{abstract}
Shear-thinning viscosity is a non-Newtonian behaviour that active particles often encounter in biological fluids such as blood and mucus. The fundamental question of how this ubiquitous non-Newtonian rheology affects the propulsion of active particles has attracted substantial interest. In particular, spherical Janus particles driven by self-diffusiophresis, a major physico-chemical propulsion mechanism of synthetic active particles, were shown to always swim slower in a shear-thinning fluid than in a Newtonian fluid. In this work, we move beyond the spherical limit to examine the effect of particle eccentricity on self-diffusiophoretic propulsion in a shear-thinning fluid. We use a combination of asymptotic analysis and numerical simulations to show that shear-thinning rheology can enhance self-diffusiophoretic propulsion of a spheroidal particle, in stark contrast to previous findings for the spherical case. A systematic characterization of the dependence of the propulsion speed on the particle's active surface coverage has also uncovered an intriguing feature associated with the propulsion speeds of a pair of complementarily coated particles not previously reported. Symmetry arguments are presented to elucidate how this new feature emerges as a combined effect of anisotropy of the spheroidal geometry and nonlinearity in fluid rheology. 
\end{abstract}

\begin{keywords}
Authors should not enter keywords on the manuscript, as these must be chosen by the author during the online submission process and will then be added during the typesetting process.
\end{keywords}

\section{Introduction}
Due to their small sizes, swimming microorganisms such as bacteria and spermatozoa live in a low-Reynolds-number world, where viscous forces dominate inertial forces. They use a variety of strategies to overcome the challenge of generating self-propulsion at low Reynolds number \citep{purcell1977life}. Extensive studies have elucidated the hydrodynamics of these biological propulsion mechanisms and shed light on their profound roles in various biological processes \citep{Fauci06,lauga2009hydrodynamics}. In recent decades, there are also growing interests in developing synthetic active particles that can self-propel like living microorganisms for biomedical and microfluidic applications, including self-assembly \citep{schwarz2012phase, wensink2014controlling}, drug delivery \citep{gao2014synthetic}, and motion-based microsensing \citep{kagan2009chemical}. Some synthetic active particle designs are inspired by biological systems, such as artificial helical propellers \citep{Zhang2009b, Ghosh2009}, which mimic the helical structure of bacterial flagella \citep{Lauga2016_Annu}. Other novel designs exploit different physical or physico-chemical mechanisms to achieve self-propulsion \citep{schweitzer2003brownian, bechinger2016active, patteson2016active,moran2017phoretic}.

In particular, a major class of synthetic active particles converts chemical energy into motility by asymmetric chemical reactions on the particle surface.   A variety of novel synthetic active colloids has been developed \citep{patino2018fundamental, buttinoni2012active, zhou2018photochemically}. For instance, microspheres half-coated in platinum, also known as Janus particles, can self-propel via catalytic decomposition of hydrogen peroxide on the platinum-coated surface \citep{Howse2007, sanchez2015chemically}. While the exact mechanism underlying the resulting motion is still under debate \citep{Brown2014,Ebbens_2014,Eloul2020}, it has been hypothesised that the motion is diffusiophoretic as a result of the gradients of molecular oxygen produced by the catalytic decomposition on the half-coated surface \citep{golestanian2005propulsion, golestanian2007designing, moran2017phoretic}. Since the solute concentration gradient is self-generated, the motion of these active particles is also referred to as self-diffusiophoresis. To model the self-diffusiophoretic motion, a common approach is to separate the fluid domain into outer (the bulk fluid) and inner (the interaction layer) regions, where the short-range solute-particle interaction is assumed to be confined in the interaction layer \citep{anderson1989colloid, Julicher2009}. When the interaction layer is thin relative to particle size, the phoretic effects can be represented by a distribution of effective slip velocities at the particle surface, analogous to the squirmer model \citep{lighthill1952squirming, blake1971spherical,Pedley16} proposed for swimming ciliates such as \textit{Paramecium} and \textit{Volvox}. While the slip velocity in the squirmer model is determined by the beating motion of short cilia covering the cell, the slip velocity of a self-diffusiophoretic particle is proportional to the solute concentration gradient and phoretic mobility calculated from the interaction potential in the interaction layer \citep{anderson1989colloid,Julicher2009}.   As a remark, recent studies have indicated that the standard self-diffusiophoretic framework described may become ineffective when the reactive species are charged \citep{brown2017ionic, de2020self,asmolov2022self}. 

Extensive studies have elucidated various interesting features of self-diffusiophoretic motion in a Newtonian fluid \citep{moran2017phoretic}. However, most biological fluids such as blood and mucus display non-Newtonian (complex) rheological behaviors including viscoelasticity and shear-thinning viscosity \citep{hwang1969rheological, baskurt2003blood}. Since these synthetic active particles will invariably encounter biological fluids in their biomedical applications, a fundamental question is how different non-Newtonian rheological behaviors impact the propulsion of these active particles \citep{patteson2016active}.  While many previous theoretical and experimental studies focused on swimming in viscoelastic fluids \citep{sznitman2014locomotion, elfring2015theory,   li2021microswimming, spagnolie2023swimming, natale2017autophoretic, de2015locomotion, zottl2019enhanced, saad2019diffusiophoresis, bechinger2016active}, recent studies have begun to address the effect of shear-thinning viscosity \citep{montenegro2013physics,velez2013waving,gagnon2014undulatory,li2015undulatory,park2016efficient,gomez2017helical}. A shear-thinning fluid loses its viscosity with applied shear due to changes in its microstructure. Such a non-Newtonian behaviour was found to impact the propulsion of various low-Reynolds-number swimmers in qualitatively different manners \citep{datt2015squirming,datt2017active,van2022effect,qu2020effects,demir2020nonlocal,qin2021propulsion}. In particular, \cite{datt2015squirming} considered a general spherical squirmer model in a shear-thinning fluid and demonstrated how shear-thinning rheology can both enhance and hinder its propulsion, depending on specific details of the slip velocity. Interestingly, in a latter study \citep{datt2017active}, spherical self-diffusiophoretic particles were found to always swim slower in a shear-thinning fluid than in a Newtonian fluid for any level of active surface coverage. This also prompts the question to what extent the conclusion of hindered swimming continues to hold for non-spherical self-diffusiophoretic particles.

Swimmers with non-spherical shapes are commonly found in both nature and engineered systems. For instance, ciliates such as \textit{Paramecium} and \textit{Tetrahymena} have approximately prolate spheroidal body shapes. \cite{keller1977porous} considered a spheroidal squirmer model, which was extended by later studies to probe the effect of geometrical shape upon ciliary locomotion \citep{ishimoto2013squirmer,theers2016modeling,poehnl2020axisymmetric}. Furthermore, synthetic active particles of non-spherical shapes, including prolate spheroids and general slender bodies,  were also fabricated and studied experimentally and theoretically \citep{champion2006role, champion2007making, glotzer2007anisotropy, shemi2018self,poehnl2020axisymmetric,zhu2023self, katsamba2022chemically,poehnl2021phoretic,yariv2019self}. In particular, \cite{poehnl2020axisymmetric} analysed the self-diffusiophertic motion of spheroidal particles in a Newtonian fluid. However, much less is known about these spheroidal active particles in non-Newtonian fluids. A recent study has suggested that shear-thinning rheology can indeed enhance the propulsion of a squirming spheroid \citep{van2022effect}. However, it remains unclear whether or not a spheroidal self-diffusiophoretic particle can swim faster in a shear-thinning fluid than in a Newtonian fluid, which was shown impossible for the spherical case \citep{datt2017active}. In this work, we fill in this knowledge gap by analysing the self-diffusiophoretic motion of a spheroidal particle in a shear-thinning fluid. We use asymptotic analysis and numerical simulations to reveal how shear-thinning viscosity impacts the propulsion speed of a prolate spheroidal self-diffusiophoretic particle with different eccentricities and levels of active surface coverage. Our results have uncovered some propulsion behaviours not observed in the spherical case and we present symmetry considerations to help elucidate the emergence of these new features as a combined effect of particle anisotropy and nonlinear fluid rheology.

\section{Problem formulation}

\subsection{Geometrical setup}

%===========================================================
\begin{figure*}
	\centering
	\includegraphics[scale=0.8]{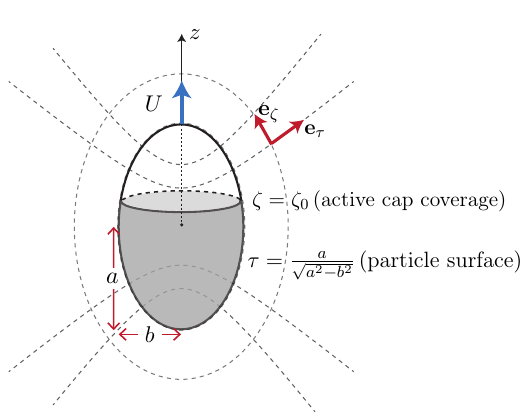}
	\caption{\footnotesize Geometric configuration of a spheroidal Janus particle. The model is presented in prolate spheroidal coordinates $(\tau,\zeta,\phi)$. The coordinate grid is indicated by dashed lines, and the basis vectors are denoted  $\e_\tau$ and $\e_\zeta$. The active cap of the particle, depicted in grey, spans from $\zeta = -1$ to $\zeta_0$. The rest of the surface is inert.  } 
	\label{fig:geo}
\end{figure*} 
%===============================================================
% GEOMETRY
We examine a prolate spheroidal particle characterised by a major axis, $a$, and a minor axis, $b$, as illustrated in figure \ref{fig:geo}.  The prolate spheroidal coordinates $(\tau, \zeta, \phi)$, where $\tau \in [1,\infty)$, $\zeta\in [-1,1]$, and $\phi\in[0, 2\pi)$, are employed in this work. The prolate spheroidal coordinates can be related to the cylindrical coordinates $(r, z, \phi)$ as 
\begin{align}\label{eq:sphe_co_trans}
	r &=\cf \sqrt{\tau^2-1} \sqrt{1-\zeta^2} ,\quad z = \cf \tau\zeta,
\end{align}
where $r^2=x^2+y^2$, and $ \cf = \sqrt{a^2-b^2}$. The surface of the spheroidal particle is given by the equation
\begin{align}
	\frac{z^2}{a^2}+\frac{r^2}{b^2} = 1,
\end{align}
which translates to $r = b\sqrt{1-\zeta^2}$ and $z = a\zeta$. Comparing with  \eqref{eq:sphe_co_trans}, the spheroidal particle surface can be simply represented  by
\begin{align}
	\tau =\tau_0 = 1/e,
\end{align}
where $e = \cf/a$ is the eccentricity. The basis vectors in the prolate spheroidal coordinates, represented as $(\e_\tau,\e_\zeta,\e_\phi)$, are related to the basis vectors in cylindrical coordinates, denoted as $(\e_r,\e_z,\e_\phi)$, in the following manner,
\begin{align}
	\begin{split}
	\e_\tau = \frac{\cf\tau}{h_\zeta} \e_r+\frac{\cf\zeta}{h_\tau}\e_z, \quad \e_\zeta = -\frac{\cf \zeta}{h_\tau}\e_r+\frac{\cf \tau}{h_\zeta}\e_z.
	\end{split}
\end{align}
 The metric coefficients for the prolate spheroidal coordinates are given by
\begin{align}\label{eq:metric}
	h_\tau = \frac{\cf \sqrt{\tau^2-\zeta^2}}{\sqrt{\tau^2-1}},\quad h_\zeta = \frac{ \cf \sqrt{\tau^2-\zeta^2}} {\sqrt{1-\zeta^2}} ,\quad  h_{\phi} =  \cf \sqrt{\tau^2-1}\sqrt{1-\zeta^2}.
\end{align}
 On the surface of the prolate spheroidal particle,  the unit normal vector pointing outwards is given by $\n = \e_\tau$, and the unit tangent vector pointing upwards is given by $\t  = \e_\zeta$ as illustrated in figure \ref{fig:geo}.

\subsection{Governing equations and boundary conditions}
We treat the problem within the continuum framework of self-diffusiophoretic propulsion \citep{golestanian2007designing,michelin2014phoretic}, where the particle interacts with a solute species of local concentration $C$. Here, we consider an axisymmetric Janus spheroidal particle with chemically active and inert compartments, with the polar position $\zeta_0$ specifying the active surface coverage as illustrated in figure \ref{fig:geo}. On the active portion of the particle surface ($\tau = \tau_0, \ \zeta\leq\zeta_0$), we assume that the solute is emitted/absorbed with a fixed-flux characterised by the activity $A$,
\begin{align}\label{eq:Cflux}
	D\n\cdot\nabla C = -A,
\end{align}
where $D$ is the diffusivity, $A>0$ corresponds to solute emission, and $A<0$ corresponds to the solute absorption. The activity becomes zero ($A=0$) on the inert portion of the particle surface ($\tau = \tau_0, \ \zeta>\zeta_0$). Under the assumption of a thin interaction layer \citep{golestanian2005propulsion, golestanian2007designing, michelin2014phoretic, datt2017active}, the effective slip velocity at the surface of the particle,
\begin{align}
	\u_s &= M (\I-\n\n)\cdot\grad C, \label{eqn:PhoreticSlip}
\end{align}
is proportional to the tangential concentration gradients and the phoretic mobility $M$ determined by the interaction potential profile \citep{anderson1989colloid, michelin2014phoretic}. In general, when the interactions are attractive, $M<0$ and the slip velocity is opposite to the concentration gradients; when the interactions are repulsive, $M>0$ and the slip velocity is along the concentration gradients. In this work, we present results for the case where $M>0$ and $A>0$ without loss of generality. By symmetry and linearity, a flipping of the sign of $M$ or $A$ only inverts the direction of swimming velocity in the results presented below.

In the bulk fluid, the solute concentration is governed by an advection-diffusion equation
\begin{align}
	\frac{\partial C}{\partial t}+\u\cdot\grad C = D\grad^2 C,
\end{align}
where $\u$ is the velocity of the flow, and the solute concentration in the far-field is denoted by $C_\infty$. 
In the inertialess regime, the flow generated by the phoretic slip velocity is governed by the momentum and continuity equations, respectively, as
\begin{align}\label{eq:stokes}
	\grad\cdot\bsigma &= \mathbf{0},\quad \grad\cdot\u = 0,
\end{align}
where $\bsigma = -p \boldsymbol{I} + \boldsymbol{T}$, $p$ is the pressure, $\boldsymbol{I}$ is the identity tensor, and $\boldsymbol{T}$ is the deviatoric stress tensor. The boundary condition for the velocity field on the particle surface in the laboratory frame is given by
\begin{align}
      	\u(\tau = \tau_0) &= \u_s(\zeta)+\boldsymbol{U},
\end{align}
where $\u_s$ is the phoretic slip velocity given by (\ref{eqn:PhoreticSlip}), and $\boldsymbol{U} = U \mathbf{e}_z$ is the unknown propulsion velocity, which occurs in the $z$-direction by axisymmetry. The flow decays to zero in the far field, $\u(\tau\to\infty) = \mathbf{0}$. The system of equations is closed by enforcing the force-free condition on the particle, 
\begin{align}\label{eq:fb}
	\int_S \n\cdot \bsigma\; {\rm d}S=  \mathbf{0},
\end{align}
where $S$ denotes the particle surface.

\subsection{Shear-thinning rheology}
To probe the effect of shear-thinning rheology on the self-diffusiophoretic motion, we consider here the Carreau constitutive model \citep{bird1987dynamics}, which has been shown effective in capturing the shear-thinning viscosity of different biological fluids \citep{velez2013waving}. In the Carreau model, the deviatoric stress is given by 
\begin{align}\label{eq:TD}
	\T =\left[ \mu_\infty+(\mu_0-\mu_\infty)\left(1+\frac{1}{2}\lambda^2\dbgamma : \dbgamma \right)^{\frac{n-1}{2}}\right]\dbgamma;
\end{align}
here $\mu_0$ and $\mu_\infty$ represent, respectively, the viscosities when the shear rate is zero and infinite, $1/\lambda$ characterizes the critical shear rate at which the non-Newtonian behaviour becomes significant, and $\dbgamma = \nabla \u+ \lp \nabla \u \rp^T$ is the strain rate tensor. For low and high shear rates (relative to the critical shear rate), the fluid tends to behave as a Newtonian fluid with viscosity, respectively, $\mu_0$ and $\mu_\infty$. In the intermediate regime, the fluid displays a power-fluid behaviour, with the index  $n < 1$ characterising the degree of shear-thinning.

\subsection{Non-dimensionalisation}

We non-dimensionalise the problem by scaling lengths with $a$,  velocities with $MA/D$, stresses by $\mu_0 MA/Da$, and the solute concentration with $Aa/D$. Hereafter we consider only dimensionless quantities and use the same symbols as their dimensional counterparts for convenience. 

We denote the solute concentration relative to the far-field solute concentration as $c=C-C_\infty$, which satisfies the dimensionless advection-diffusion equation,
\begin{align}
	\Pen \left(\frac{\partial c}{\partial t}+\u\cdot\grad c\right)= \grad^2 c.
\end{align}
Here, the P\'eclet number $\Pen = MAa/D^2$ characterises the relative importance of advective to diffusive transport of the solute. We assume the diffusivity is high enough and neglect the alteration in solute distribution caused by the flow from phoretic effects,  the solute concentration becomes harmonic,
\begin{align}\label{eq:cnd}
    \grad^2 c= 0.
\end{align}
The dimensionless boundary condition on the active portion of the particle surface ($\tau = \tau_0, \ \zeta\leq\zeta_0$) is given by 
\begin{align}\label{eq:cbcnda}
	\n\cdot\grad c = -1, 
\end{align}
whereas that on the inert portion ($\tau = \tau_0, \ \zeta>\zeta_0$) is simply 
\begin{align}\label{eq:cbcndb}
	\n\cdot\grad c = 0.
\end{align}
The relative solute concentration decays to zero at infinity, 
\begin{align} \label{eqn:cbc}
c(\tau \rightarrow \infty) = 0.
\end{align}

Given that the solute concentration is decoupled, the governing equations for the fluid align with those presented in (\ref{eq:stokes}). The Carreau constitutive equation is rendered dimensionless as
\begin{align}\label{eq:TND}
	\T =\dbgamma+ (1-\beta)\left[ -1 +\left(1+\frac{1}{2}\Can^2\dbgamma : \dbgamma \right)^{\frac{n-1}{2}}\right]\dbgamma,
\end{align}
where $\beta = \mu_\infty/\mu_0$, the viscosity ratio, and $\Can= \lambda MA/aD$, the Carreau number, which compares the characteristic shear rate $MA/aD$ to the critical shear rate $1/\lambda$. 

In the laboratory frame, the dimensionless boundary condition for the velocity field on the particle surface is given by $\u(\tau = \tau_0) = \boldsymbol{u}_s + U \e_z$, where the slip velocity in dimensionless form reads
\begin{align}\label{eq:usnd}
	\boldsymbol{u}_s = (\I-\n\n)\cdot\grad c,
\end{align}
and the flow decays to zero in the far-field, $\u(\tau\to\infty) = \mathbf{0}$. In the following calculations, we determine the unknown propulsion speed $U$ of the spheroidal self-diffusiophretic particle in a shear-thinning fluid.
\section{Asymptotic analysis and numerical simulations}

\subsection{Asymptotic analysis} \label{sec:Asym}
The solute concentration can be obtained by solving the Laplace equation (\ref{eq:cnd}) with boundary conditions  (\ref{eq:cbcnda})--(\ref{eqn:cbc}) in the prolate spheroidal coordinates. An analytical solution in form of a series is given by \citep{popescu2010phoretic}
\begin{align}\label{eq:C}
	c(\tau,\zeta) = \sum_{n=0}^\infty \rho_nQ_n(\tau)P_n(\zeta),
\end{align}
where $P_n(\zeta)$ and $Q_n(\tau)$ are, respectively, the Legendre functions of the first and the second kinds.  As $\tau\geq\tau_0>1$, the Legredre functions of the the second kind vanish when $\tau\to\infty$, satisfying the far-field boundary condition for the relative concentration, (\ref{eqn:cbc}). %The expression of the Legendre functions of the second kind are given in Appendix B.
By substituting the solution \eqref{eq:C} into the boundary conditions at the particle surface, (\ref{eq:cbcnda})--(\ref{eq:cbcndb}), and employing the orthogonality of the Legendre functions, the coefficients $\rho_n$ in the series solution are determined as
\begin{align}
	\rho_n(\tau_0,\zeta_0) = -\frac{2n+1}{2}\frac{1}{Q'_n(\tau_0)\tau_0\sqrt{\tau_0^2-1}}\int_{-1}^{\zeta_0} \sqrt{\tau_0^2-\zeta^2}P_n(\zeta) \ \text{d}\zeta.
	\label{eq:concentration coefficients}
\end{align}
By employing the solution \eqref{eq:C} in \eqref{eq:usnd}, the resulting phoretic slip velocity at the particle surface, $\mathbf{u}_s(\zeta) = u_s \e_\zeta$, is given by  
\begin{align}
	u_s(\zeta)	= \tau_0\sum_{n=0}^\infty B_n \frac{P^1_n(\zeta)}{\sqrt{\tau_0^2-\zeta^2}}, \label{eq:ussol}
\end{align}
where the phoretic modes are given by 
\begin{align}
    	B_n = -\rho_n(\tau_0,\zeta_0) Q_n(\tau_0),
	\label{eq:modes}
\end{align}
and $P_n^1$ is the associated Legendre function with order 1.

We perform an asymptotic analysis in the weakly non-Newtonian regime where the deviation of the viscosity ratio from unity, $\epsilon = 1-\beta$, is small.  We expand the physical quantities in powers of $\epsilon$ as
\begin{align}\label{eq:expan}
	\{\u, \dbgamma, \boldsymbol{\sigma}, p, \T,U\} = \{\u_0, \dbgamma_0,\boldsymbol{\sigma}_0, p_0, \T_0,U_0\} +\epsilon \{\u_1, \dbgamma_1,\boldsymbol{\sigma}_1, p_1, \T_1,U_1\}+O(\epsilon^2).
\end{align}
The zeroth-order problem corresponds to the Newtonian problem, where $\boldsymbol{\sigma}_0 =-p_0\I+ \dbgamma_0$, and $\dbgamma_0 = \grad\u_0+(\grad\u_0)^T$. For boundary conditions, we have $\u_0(\tau = \tau_0) =U_0\e_z+ u_s \e_\zeta$ on the particle surface and $\u_0(\tau\to \infty) = \mathbf{0}$ in the far-field, where $u_s$ is given in \eqref{eq:ussol}. The flow field $\u_0$ and propulsion speed $U_0$ of this zeroth-order, Newtonian problem were obtained in previous works \citep{Leshansky_2007, Lauga2016,popescu2010phoretic, poehnl2020axisymmetric}, which we summarise in Appendix \ref{sec:appenA}.

We consider the first-order non-Newtonian correction to the Newtonian problem. To the order of $\epsilon$, the flow satisfies 
\begin{align}
\grad\cdot\bsigma_1 &= \mathbf{0}, \\
\grad\cdot\u_1 &= 0,
\end{align}
where $\bsigma_1 = -p_1 \I+\T_1$ and the stress tensor $\T_1 = \dbgamma_1 + \mathsfbi{A}$, with 
 \begin{align}
  \mathsfbi{A} = \left[ -1 +\left(1+\frac{1}{2}\Can^2\dbgamma_0 : \dbgamma_0 \right)^{\frac{n-1}{2}}\right]\dbgamma_0. 
  \end{align}
For boundary conditions, we have the first correction to the Newtonian propulsion velocity, $\u_1(\tau = \tau_0) =\U_1 = U_1\e_z$, on the particle surface, and $\u_1(\tau\to \infty) = \mathbf{0}$ in the far-field. To obtain the propulsion speed $U_1$, we bypass detailed calculations of the flow via a reciprocal theorem approach \citep{lauga2014locomotion}. By considering an auxiliary Stokes flow due to a prolate spheroid of the same geometry translating at a velocity $\hat{\U}$, where the velocity $\hat{\u}$ and stress $\hat{\bsigma}$ fields satisfy $\grad\cdot \hat{\bsigma} = \mathbf{0}$ and $\grad\cdot\hat{\u}=0$, one can form  the relation
\begin{align}
\hat{\u}\cdot(\grad\cdot\bsigma_1) = \u_1\cdot(\grad\cdot\hat{\bsigma}) = 0.
\end{align}
By integrating the relation over the fluid volume  $V$ exterior to the particle surface $S$ and applying the divergence theorem, one can obtain
\begin{align}
\int_S{\boldsymbol{n}\cdot\hat{{\boldsymbol{\sigma}}}\cdot{\boldsymbol{u}}_1} \ \text{d}S - \int_S{\boldsymbol{n}\cdot{\boldsymbol{\sigma}}_1 \cdot{\hat{\boldsymbol{u}}}} \ \text{d}S=\int_V{\boldsymbol{\sigma}_1:\nabla\hat{\boldsymbol{u}}}\ \text{d}V-\int_V{\hat{\boldsymbol{\sigma}}:\nabla\boldsymbol{u}_1} \ \text{d}V.
    \label{recirpcoal gen 2a}
\end{align}
We note that due to the force-free condition at $O(\epsilon)$, $\int_S{\boldsymbol{n}\cdot{\boldsymbol{\sigma}}_1 } \ \text{d}S = \mathbf{0}$, the second integral on the left hand side of \eqref{recirpcoal gen 2a} is given by  $\int_S{\boldsymbol{n}\cdot{\boldsymbol{\sigma}}_1 \cdot{\hat{\boldsymbol{u}}}} \ \text{d}S = (\int_S{\boldsymbol{n}\cdot{\boldsymbol{\sigma}}_1 } \ \text{d}S)\cdot \hat{\U} = 0$. Upon substituting the constitutive equations for $\hat{\boldsymbol{\sigma}}$ and $\boldsymbol{\sigma}_1$ and applying the boundary condition $\boldsymbol{u}_1 = \U_1$ on $S$, \eqref{recirpcoal gen 2a} simplifies to
\begin{align}
\hat{\boldsymbol{F}} \cdot \boldsymbol{U}_1  =\int_V{\mathsfbi{A}:\nabla\hat{\boldsymbol{u}}}\ \text{d}V,
    \label{swimming speed correction integral theorem}
\end{align}
where $\hat{\boldsymbol{F}} = \int_S \n\cdot \hat{\bsigma} \ \text{d}S =-8\pi \tau_0^{-1}\left[(\tau_0^2+1)\coth^{-1}\tau_0-\tau_0\right]^{-1}\e_z$ is the drag on the translating prolate spheroid in the auxiliary problem. Therefore, the first-order correction to the phoretic speed is given in terms of a volume integral in prolate spheroidal coordinates as
\begin{align}\label{eq:U1}
	U_1 =-\frac{(\tau_0^2+1)\coth^{-1}\tau_0-\tau_0}{4 \tau_0^2}\int_{\tau_0}^\infty \int_{-1}^1 \left(\mathsfbi{A}:\nabla \hat{\boldsymbol{u}}\right)(\tau^2-\zeta^2) \;\text{d}\zeta \text{d}\tau,
\end{align}
which can be evaluated with quadrature.

\begin{figure}
	\centering
	\includegraphics[scale=1]{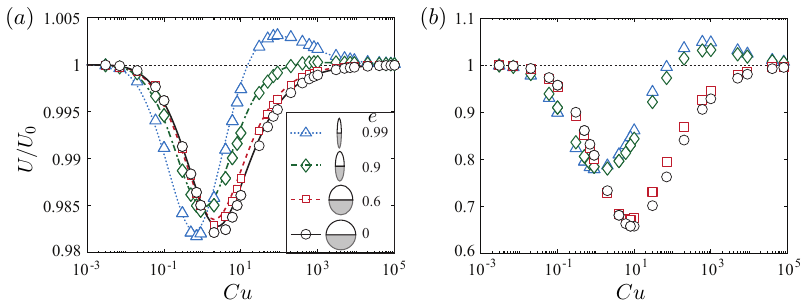}
	\caption{$(a)$ Swimming speed of a spheroidal Janus particle $U$ in a shear-thinning fluid relative to its corresponding Newtonian value $U_0$ as a function of the Carreau number for different values of eccentricity $e$ when the shear-thinning effect is weak ($\beta= 0.9$). The asymptotic results in the small $\epsilon = 1-\beta$ limit (lines) agree well with numerical simulations (symbols). For large eccentricies (e.g.  $e = 0.9$ and 0.99), the Janus particle can swim faster in a shear-thinning fluid than in a Newtonain fluid. $(b)$ Numerical results for strong shear-thinning effect ($\beta= 0.1$), the qualitative behaviours remain the same, the speed variations are substantially larger. In both $(a,b)$, the active coverage of the particle $\zeta_0 = 0$ and the shear-thinning power law index $n = 0.25$.} 
	\label{fig:velocity}
\end{figure} 

\subsection{Numerical simulation} \label{sec:Numerical}
To extend the results beyond the weakly non-Newtonian regime considered
in the asymptotic analysis in \S \ref{sec:Asym}, we develop numerical simulations based on the finite element method (FEM) using the partial differential equation (PDE) module of the commercial package COMSOL to perform fully coupled simulations of the momentum and continuity equations \eqref{eq:stokes} with the Carreau-Yasuda constitutive equation \eqref{eq:TND}, and the solute transport equation \eqref{eq:cnd}. We use an axisymmetric computational domain with a dimensionless radius of $500$ to simulate the self-propulsion of the Janus particle in an unbounded fluid. A sufficiently large domain size is important to guarantee accuracy due to the slow spatial decay of flows at low Reynolds numbers. The Janus particle is modelled as a half-spheroid whose major axis coincides with the axis of symmetry. The simulations are performed in a reference frame that is co-moving with the particle, and the far-field velocity is equal to the negative swimming velocity determined by the force-free condition \eqref{eq:fb}. 
The computational domain is discretised by approximately $100000-127000$ triangular elements, and the mesh is locally refined near the particle to properly resolve the spatial variation of the viscosity. 
Taylor-Hood and quadratic Lagrange elements are adopted to discretise the flow field ($\u, p$) and the concentration field $c$, respectively. 
It is important to note that, theoretically, there exists a discontinuous alteration in surface activity between the active and inert compartments of the Janus particle. However, when modelled numerically, this abrupt transition can cause significant numerical errors, particularly at lower $\Can$ values. To alleviate the numerical errors, we introduce a minor smoothing transition, dependent on the mesh size, to the surface activity in the vicinity of the discontinuity.

In addition to comparing with the asymptotic results in this work, we have validated our numerical implementation against previous results for a spherical Janus particle in a shear-thinning fluid \citep{datt2017active} and a spheroidal Janus particle in a Newtonian fluid \citep{popescu2010phoretic}; see Appendix B for more details.

\section{Results and discussion}
%===========================================================
\begin{figure*}
	\centering
	\includegraphics[scale=0.55]{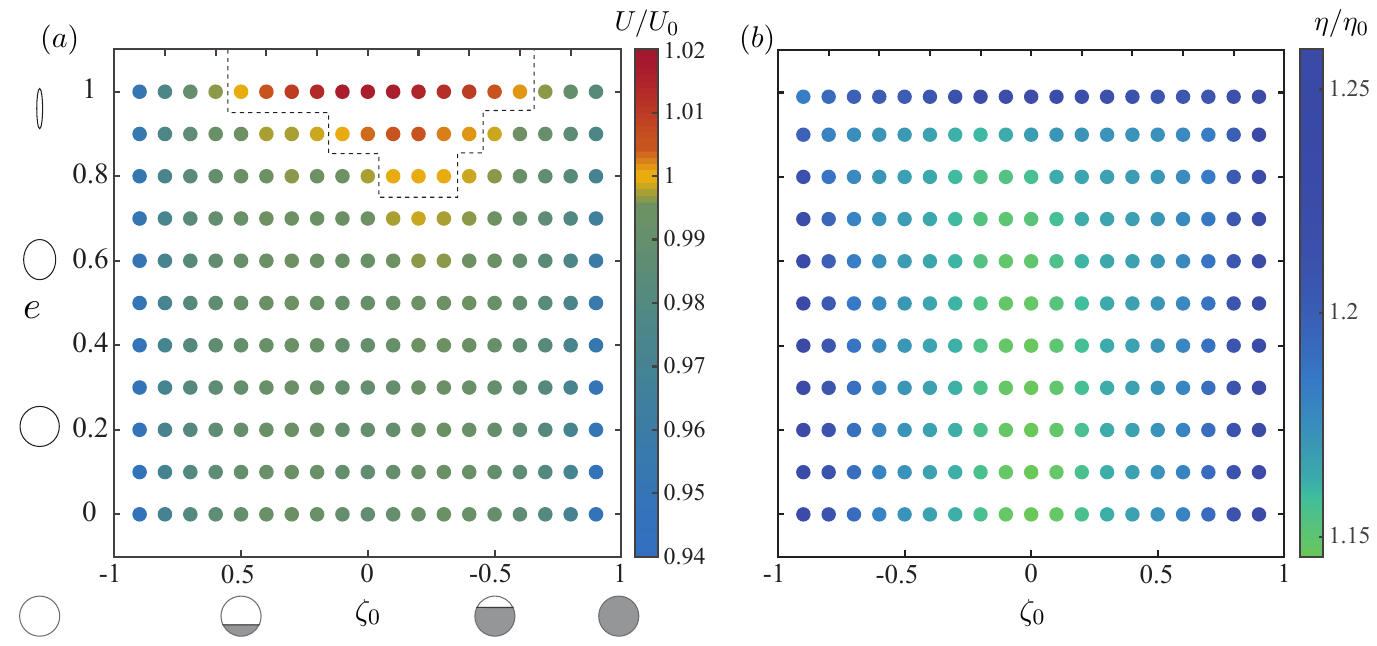}
	\caption{\footnotesize $(a)$ Swimming speed of a spheroidal Janus particle in a shear-thinning fluid with different values of eccentricity $e$ and active coverage $\zeta_0$. The dashed line indicates the particles of which the swimming speed is enhanced by the shear-thinning effect. $(b)$ Relative swimming efficiency of a spheroidal Janus particle  with different values of eccentricity and active coverage.  For all data points,  $\Can = 20000$,  $\beta = 0.1$ and $n = 0.25$.  
} 
	\label{fig:phase}
\end{figure*} 
%===============================================================

\subsection{Effect of particle eccentricity on self-diffusiophoresis in a shear-thinning fluid} \label{sec:eccentricity}

In a Newtonian fluid, the dependence of the self-diffusiophoretic propulsion speed on the particle geometry and catalyst coverage was examined in detail by previous works \citep{popescu2010phoretic, poehnl2020axisymmetric}. Here we investigate how shear-thinning rheology impacts the propulsion speeds ($U$) relative to their corresponding Newtonian values ($U_0$). The special case of a spherical Janus particle was examined by \cite{datt2017active} and shown to always swim slower in a shear-thinning fluid than in a Newtonian fluid across a wide range of $\Can$. In figure \ref{fig:velocity}$(a)$, we reproduce these results by setting the eccentricity to be zero ($e=0$, black solid line and black circles): the spherical Janus particle displays reduced propulsion speed ($U/U_0 < 1$) as $\Can$ increases from zero, reaching a local minimum when $\Can$ is around $\mathcal{O}(1)$, before approaching the Newtonian value again when $\Can$ becomes exceedingly large. We employ the spherical case as a benchmark to probe the effect of particle geometry by varying the eccentricity from $e=0$ to $e=0.99$. From spherical to moderately spheroidal particles (e.g., $e=0.6$), the increased eccentricity does not affect the qualitative features of the speed dependence on $\Can$. 

However, for more slender spheroidal particles (e.g., $e=0.99$), our results reveal that a self-diffusiophoretic particle can also swim faster in a shear-thinning fluid than in a Newtonian fluid (blue dotted lines and blue upward triangles), which was shown impossible for a spherical particle (black solid line and black circles) \citep{datt2017active}. These new behaviors are predicted by both the asymptotic results by the reciprocal theorem (lines) and results by numerical simulations (symbols) in the weakly shear-thinning regime ($\beta=0.9$), which display excellent agreements as shown in figure \ref{fig:velocity}$(a)$. We verify that these new features continue to exist beyond the weakly non-Newtonian regime by considering a small viscosity ratio ($\beta=0.1$) in figure \ref{fig:velocity}$(b)$, where we observe the same qualitative behaviours but with greater magnitudes of speed enhancement and reduction at different $\Can$.

%===============================================================
\begin{figure}
	\centering
	\includegraphics[scale=0.9]{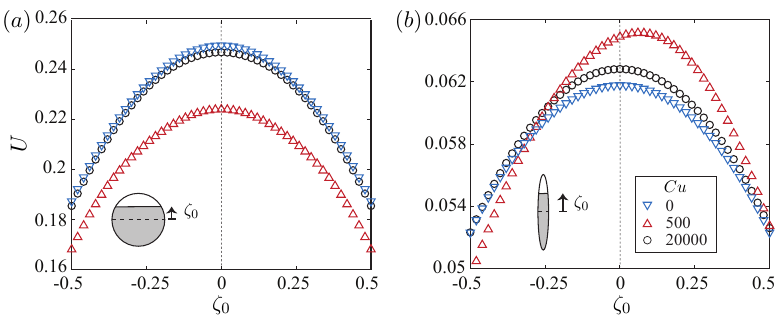}
	\caption{Swimming speed of $(a)$ a spherical $(e=0)$ and $(b)$ a spheroidal $(e=0.99)$ Janus particle  as a function of $\zeta_0$  with  $\beta = 0.1$. Three fluids are considered: $\Can = 0$ (blue downward-pointing triangle, Newtonian fluid), 
   500 (red upward-pointing triangle), 20000 (black circle).}
	\label{fig:speedmax}
\end{figure} 
%===============================================================

\subsection{Effect of active surface coverage on self-diffusiophoresis in a shear-thinning fluid}

We focus in \S \ref{sec:eccentricity} on Janus particles with half active surface coverage ($\zeta_0 = 0$),  which was shown to maximise the self-diffusiphoretic propulsion speed of spherical and spheroidal particles in a Newtonian fluid. Here, we examine whether this feature remains the same or not when the fluid displays shear-thinning rheology. In figure \ref{fig:phase}$(a)$, we display the propulsion speed relative to its Newtonian value as a function of particle eccentricity and active surface coverage, which varies between $\zeta_0=-1$ (no active surface coverage) and $\zeta_0=1$ (full active surface coverage). It is observed that, regardless of the active surface coverage, the regime of enhanced propulsion ($U/U_0>1$, indicated by the dashed line in figure \ref{fig:phase}$(a)$) only occurs when the particle eccentricity goes beyond a threshold value of approximately 0.7. In addition, the enhanced propulsion occurs for a wider range of active surface coverage with increased particle eccentricity. For instance, among all the values of active surface coverage examined in figure \ref{fig:phase}$(a)$, while enhanced propulsion is observed only in about $15\%$ of the cases when $e=0.8$, the percentage increase to more than 60\% when $e=0.99$. Another interesting feature is the asymmetry in the occurrence of enhanced propulsion with respect to the active surface coverage: the regime is not symmetrically distributed around $\zeta_0$ but instead skewed towards the positive direction of $\zeta$. This observation also suggests that the specific case of half active coverage ($\zeta_0=0$), which was shown to maximise self-diffusiophertic propulsion in previous works \citep{popescu2010phoretic, poehnl2020axisymmetric, datt2017active}, may no longer be optimal for spheroidal particles in shear-thinning fluids.

In addition to propulsion speed, efficiency is another relevant performance measure of the swimming motion. Recent studies have investigated how the geometrical shape of active particles influence their efficiency of swimming in a Newtonian fluid \citep{guo2021optimal, daddi2021optimal}. Here we adopt the widely used definition of swimming efficiency introduced by \cite{Lighthill1975} for low-Reynolds-number swimmers, $\eta = \boldsymbol{F} \cdot \boldsymbol{U}/P$, to characterize the efficiency of swimming in a shear-thinning fluid. Lighthill's efficiency compares the power dissipation of the swimmer, $P = \int \n \cdot \bsigma \cdot \u \;\dtex S$, with the power required to move a particle with identical shape at the same swimming velocity $\boldsymbol{U}$ against the drag force $\boldsymbol{F}$. Our results show that, while speed enhancement occurs only in a specific domain of eccentricity and active surface coverage (figure \ref{fig:phase}$(a)$), the swimming efficiency in a shear-thinning fluid is consistently enhanced, $\eta/\eta_0>1$, relative to the corresponding swimming efficiency in a Newtonian fluid ($\eta_0$) in the entire domain shown in figure \ref{fig:phase}$(b)$. Taken together, these results reveal that self-diffusiophoretic propulsion can be enhanced both speed-wise and efficiency-wise in a shear-thinning fluid relative to the corresponding case in a Newtonian fluid.

%===========================================================
\begin{figure*}
	\centering
	\includegraphics[scale=1]{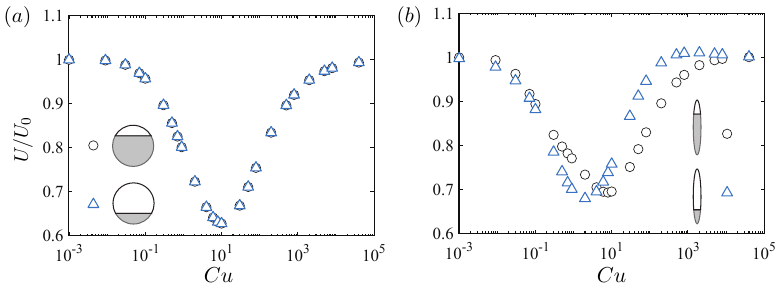}
	\caption{\footnotesize Swimming speed of $(a)$ a pair of complementarily coated spherical $(e=0)$ and $(b)$ spheroidal $(e=0.99)$ particles in a shear-thinning fluid with active coverage $\zeta_0 = \pm 0.5$.  The compementarily coated spherical swimmers are propelled with the same speed, while spheroidal swimmers break this symmetry. In both $(a,b)$, $\beta = 0.1$ and $n = 0.25$.} 
	\label{fig:asymmetry}
\end{figure*} 
%===============================================================
Next, we further examine the asymmetry observed in the enhanced propulsion speed with respect to the active surface coverage shown in figure \ref{fig:phase}$(a)$. We display in figure \ref{fig:speedmax}$(a)$ the absolute propulsion speed of a spherical and a spheroidal particles as a function of active surface coverage at different values of $\Can$. In figure \ref{fig:speedmax}$(a)$, we observe that the propulsion speed of a spherical particle is symmetric about the half surface coverage ($\zeta_0$), which maximises the speed in both Newtonian ($\Can=0$) and shear-thinning ($\Can>0$) fluids. In contrast, for a spheroidal particle with $e=0.99$  shown in figure \ref{fig:speedmax}$(b)$, while the aforementioned features still hold in the Newtonian limit ($\Can=0$, blue downward triangles), when the fluid is shear-thinning (e.g., $\Can=500$, red upward triangles) the variation of the propulsion becomes asymmetric about $\zeta_0=0$, which no longer maximises the self-diffusiophoretic propulsion speed. 
Instead, the maximum propulsion speed occurs at a positive active surface coverage ($\zeta_0>0$) as shown in figure \ref{fig:speedmax}$(b)$, depending on parameters measuring the shear-thinning effect including $\beta$ and $\Can$. The emergence of this novel feature requires the combined presence of both non-Newtonian rheology and non-spherical geometry, which we attempt to better understand via symmetry arguments presented in the next section.

\subsection{Symmetry considerations}
%===========================================================
\begin{figure*}
	\centering
	\includegraphics[scale=0.9]{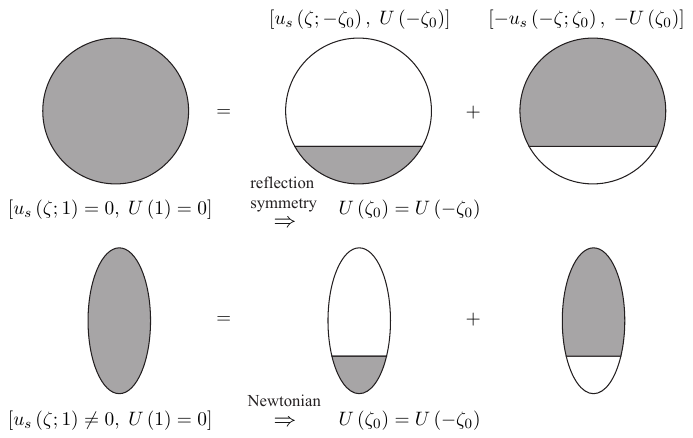}
	\caption{\footnotesize Schematics illustrating the symmetry considerations for a pair of complementarily coated particles. The phoretic slip velocity on the surface of a particle with an active coverage $\zeta_0$ is denoted as $u_s(\zeta; \zeta_0)$,  and the corresponding propulsion speed as $U (\zeta_0)$. The slip velocity on a fully-coated spherical particle is zero everywhere due to the isotropy. Consequently, the flow  induced by the particles with active region $\pm\zeta_0$ always shows a symmetry, which leads to identical speeds.   As the slip velocity on a fully-coated anisotropic particle is not zero, the flow and the slip velocity do not have the reflection symmetry, and the particle speeds are not the same in general. However, if the flow is Newtonian, the speeds are the same due to the linearity.    } 
	\label{fig:sche}
\end{figure*} 
%===============================================================

To examine this feature of symmetry breaking across the full range of $\Can$, we compare the swimming speed of two complementarily coated particles with $\zeta_0 = \pm 0.5$ for the spherical ($e = 0$, figure \ref{fig:asymmetry}$a$) and spheroidal ($e = 0.99$, figure \ref{fig:asymmetry}$b$) cases. For spherical particles, figure \ref{fig:asymmetry}($a$) shows that the swimming speed of particle with $\zeta_0 = -0.5$ (blue triangles) is identical to that with $\zeta_0 = 0.5$ (black circles) over the entire range of $\Can$, despite the latter having a significantly larger active surface coverage. On the contrary, Figure \ref{fig:asymmetry}$(b)$ demonstrates that the swimming speed of two complementarily coated spheroidal particles ($\zeta_0 = \pm 0.5$) only approach the same value when $\Can$ is exceedingly small or large, where the fluid medium becomes effectively Newtonian. At intermediate values of $\Can$, the spheroidal particle with $\zeta_0 = -0.5$ (blue triangles) generally exhibits a considerably different swimming speed compared with its complementarily coated counterpart ($\zeta_0 = 0.5$, black circles) as shown in figure \ref{fig:asymmetry}$(b)$.

\begin{figure*}
	\centering
	\includegraphics[scale=0.55]{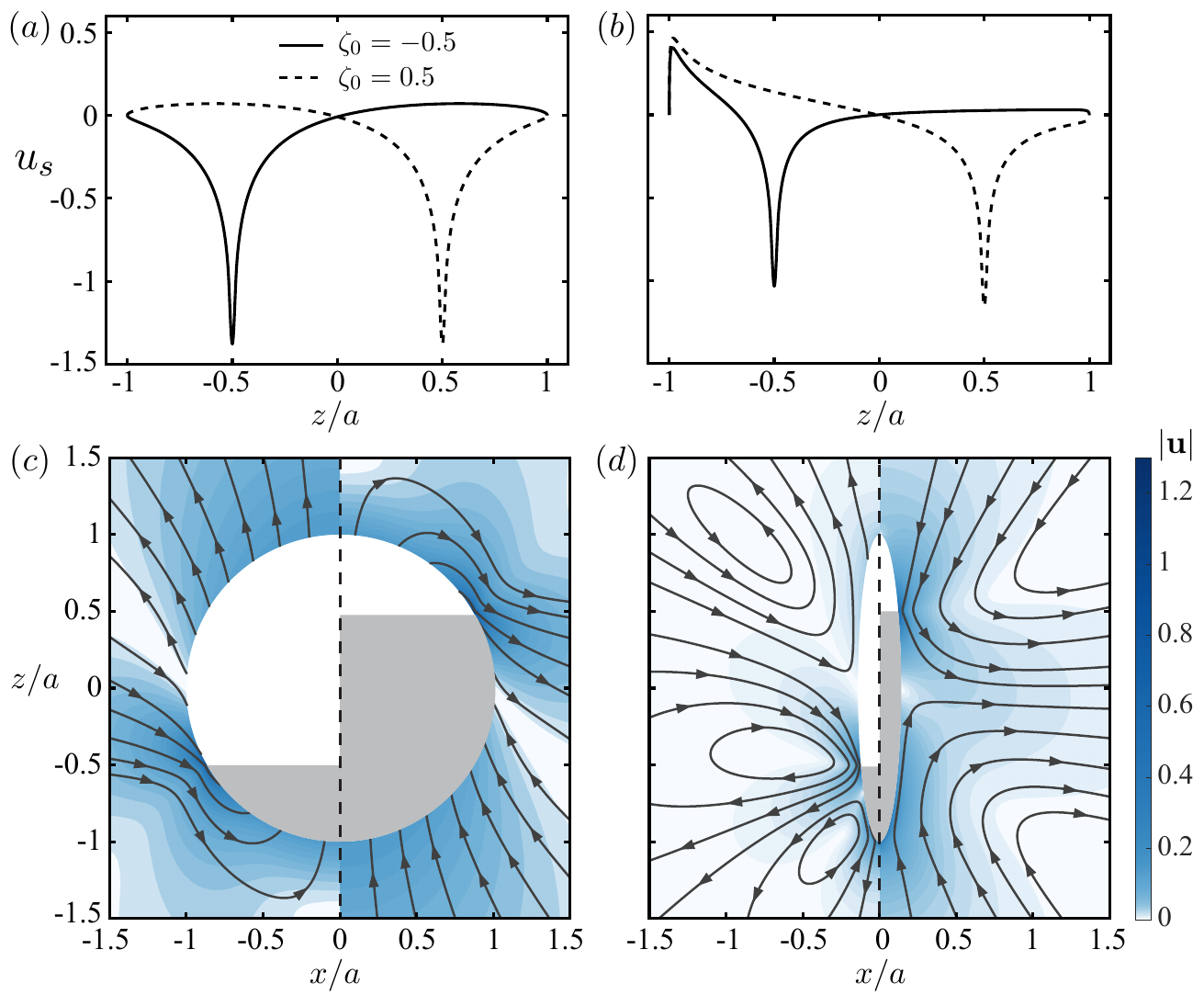}
	\caption{\footnotesize The slip velocity and the flow field around the particles with active region $\zeta_0 = \pm 0.5$,  in a shear-thinning fluid with $\Can = 1$, $\beta = 0.1$ and $n = 0.25$. The slip velocity and the flow field around the spherical swimmers has an upside-down symmetry $(a)(c)$, which does not appear in those around the spheroidal swimmers ($e = 0.99$)  $(b)(d)$. } 
	\label{fig:flow}
\end{figure*}

One may understand the above feature as a combined result of symmetry breaking and nonlinear rheology as illustrated in figure \ref{fig:sche}. We denote the phoretic slip velocity on the surface of a particle with an active coverage $\zeta_0$ as $u_s(\zeta;\zeta_0)$ and the corresponding propulsion speed as $U(\zeta_0)$. We note that in the zero-$Pe$ limit considered here, the linearity of the Laplace equation allows superposition in the solute concentration problem. Now, consider a fully coated particle, the phoretic slip velocity can be decomposed into two complementary cases, $u_s(\zeta;1) = u_s(\zeta;-\zeta_0)-u_s(-\zeta;\zeta_0)$, as shown in figure \ref{fig:sche} for a spherical and a spheroidal particles. For a fully coated spherical particle, the slip velocity is zero everywhere on the particle surface due to isotropy, $u_s(\zeta;1)=0$. This property leads to the result $u_s(\zeta;-\zeta_0) = u_s(-\zeta;\zeta_0)$, which means that the boundary conditions on two complementarily coated spherical particles become identical upon a reflection about $z=0$. This result is illustrated in figure \ref{fig:flow}$(a)$ for the slip velocity of two complementarily coated spherical particles, which consequently, upon a reflection about $z=0$, generate the same flow field  as shown in figure \ref{fig:flow}$(c)$. The identical propulsion speed of these complementarily coated particles, $U(\zeta_0)=U(-\zeta_0)$, is therefore a direct result of isotropy for spherical self-diffusiophertic particles, regardless of whether the fluid is Newtonian or non-Newtonian.

%===========================================================
%===============================================================

When particle eccentricity is introduced, the anisotropy implies that the slip velocity of a fully coated spheroidal particle does not vanish everywhere on the surface of the particle, $u_s(\zeta; 1) \neq 0$. As illustrated in figure \ref{fig:flow}$(b)$, the boundary conditions on two complementarily coated spheroidal particles therefore no longer have the reflection symmetry about $z=0$, $u_s(\zeta;-\zeta_0) \neq u_s(-\zeta;\zeta_0)$, leading to generally distinct flows surrounding the spheroidal particles as shown in figure \ref{fig:flow}$(d)$. Consequently, unlike spherical particles, one may expect two complementarily coated spheroidal particles to have distinct propulsion speeds in general. This conclusion is largely true as shown in figure \ref{fig:asymmetry}$(b)$, except for the special case when the fluid is Newtonian. For a Newtonian fluid, the linearity of the governing equations allows the superposition of the solutions associated with the pair of particles with complementary coatings to form the solution of a particle with full coating (figure \ref{fig:sche}$a$), leading to the result, $U(1) = U(\zeta_0)-U(-\zeta_0)$. Since $U(1) = 0$ for a fully coated particle, we obtain the conclusion $U(\zeta_0) = U(-\zeta_0)$ for a Newtonian fluid, which holds for both spherical and spheroidal particles, despite the absence of reflection symmetry in their slip velocities. When the fluid is non-Newtonian, the superposition described above no longer holds, allowing the propulsion speed of two complementarily coated spheroidal particles to be different.

To summarise, isotropy in spherical geometry alone guarantees that two complementarily coated particles propel with identical speeds, regardless of whether the fluid is Newtonian or not. In parallel, in a Newtonian fluid the linearity of the problem alone guarantees the same, regardless of whether the particle is spherical or not. Hence, to propel two complementarily coated particles with different speeds, both istropy and linearity need to be broken. The emergence of different speeds for a pair of complementarily coated spheroidal particles in a shear-thinning fluid reported here, therefore, serves as a specific example illustrating this general feature.

\section{Concluding remarks}

Shear-thinning viscosity is a non-Newtonian behavior that active particles often encounter in biological fluids. The investigation into how this ubiquitous non-Newtonian rheology impacts the propulsion speed of active particles has garnered considerable recent interest. In particular, previous studies have demonstrated how shear-thinning rheology slows down spherical active particles \citep{datt2015squirming, datt2017active}. A more recent investigation \citep{van2022effect} has suggested that, by tuning the geometrical shape of a squirmer, it is possible for a spheroidal squirmer to swim faster in a shear-thinning fluid than in a Newtonian fluid. In this work, we have extended the analysis by \cite{van2022effect} on the spheroidal squirmer model to self-diffusiophoretic particles, a major physico-chemical propulsion mechanism of synthetic active particles. Unlike the squirmer model, where the velocity distribution on the particle surface is prescribed, the effective slip velocity of a self-diffusiophoretic particle is determined by the solute concentration gradient and the phoretic mobility. Using asymptotic analysis to probe the weakly non-Newtonian behavior, we have demonstrated that shear-thinning viscosity can indeed enhance self-diffusiophoretic propulsion of spheroidal particles with a large particle eccentricity in a specific regime of Carreau number. This result is in stark contrast with spherical self-diffusiophoretic particles, which always swim slower in a shear-thinning fluid \citep{datt2017active}. We have also used numerical simulations to verify that the new features uncovered by the asymptotic analysis continue to hold beyond the weakly non-Newtonian regime. 

We have also systematically characterised the dependence of the self-diffusiophoretic propulsion speed on the particle's active surface coverage in a shear-thinning fluid. Previous studies showed that a pair of complementarily coated spherical or spheroidal particles always propel at the same speed in a Newtonian fluid. When the fluid becomes shear-thinning, the same propulsion speed still occurs when the complementarily coated particles are spherical in shape. However, we have found distinct propulsion speeds for two complementarily coated spheroidal particles in a shear-thinning fluid. We have also presented symmetry arguments to better understand how this new feature emerges as a combined effect of anisotropy associated with the spheroidal geometry and nonlinearity associated with the non-Newtonian rheology. Such symmetry-breaking might hint at using anisotropic active particles as a tool for probing microrheology of complex fluids.

We remark on several limitations of the current work and discuss potential directions for further investigations. First, we have neglected the effect of solute advection by considering the zero $Pe$ limit. It remains unclear how the flow modifications due to shear-thinning rheology influence solute advection and thereby the phoretic propulsion. In particular, the symmetry considerations presented in Section 4.3, which require the linearity of the Laplace equation, would no longer hold for finite $Pe$. It would therefore be interesting to probe how the nonlinearity associated with solute advection affects the symmetry breaking observed for the propulsion of complementarily coated particles. Second, we have followed previous work \citep{datt2017active} to focus on the non-Newtonian effect in the bulk fluid in this work, neglecting the influence of fluid rheology on the surface slip of a self-diffusiophretic particle, which was shown to modify the propulsion speed of a spherical Janus particle in intriguing manners \citep{choudhary2020non}. In particular, in a weakly shear-thinning fluid, the effect due to the modified slip velocity could dominate the retardation due to the bulk non-Newtonian stress, leading to the speed enhancement of a Janus spherical particle. An investigation is currently underway to extend the analysis beyond the weakly non-Newtonian regime and examine the effect of particle geometry in this more complex physical scenario, where the non-Newtonian effects on both slip and mobility of self-diffusiophoretic particles are taken into account. We also call for future efforts in developing a comprehensive physical understanding of the findings reported in this work. Last, we focus on the effect of shear-thinning viscosity here while complex biological fluids also display other non-Newtonian fluid behaviors including viscoelasticity. Future work accounting for the viscoelastic stress and its combined effects with shear-thinning rheology will shed light on how the geometric shape of self-diffusiophretic particles should be tuned to maximize their propulsion in biological fluids.\\

\noindent \textbf{Funding.} We acknowledge the supports from the National Science Foundation under grant numbers 1931292 and 2323046 (to O.S.P.), the National Natural Science Foundation of China under grant number 12372258 (to Y.M.), and the Fundamental Research Funds for the Central Universities, Peking University (to Y.M.). 
L.Z. acknowledges the partial support from the Singapore Ministry of Education Academic Research Fund Tier 2 (MOE-T2EP50122-0015). The computation of the work was performed on resources of the National Supercomputing Centre, Singapore (https://www.nscc.sg).\\

\noindent \textbf{Declaration of interests.} The authors report no conflict of interest.

\appendix

\section{Solution to the zeroth-order (Newtonian) problem}
\label{sec:appenA}
Here we summarise the solution to the zeroth-order problem, which corresponds to the self-diffusiophertic motion of a spheroidal particle in a Newtonian fluid considered in previous works \citep{Leshansky_2007, Lauga2016,popescu2010phoretic, poehnl2020axisymmetric}. In particular, we follow the approach by \cite{poehnl2020axisymmetric} here to determine the unknown velocity field $\u_0$ and propulsion velocity $\U_0$.

The unknown propulsion velocity $\U_0$ can be obtained using the Lorentz reciprocal theorem \citep{stone1996propulsion, popescu2010phoretic, poehnl2020axisymmetric}, by considering an auxiliary Stokes flow problem $(\hat{\u}, \hat{\bsigma})$ of a translating prolate spheroidal particle of the same geometry along its major axis. Via the reciprocal theorem \citep{popescu2010phoretic, poehnl2020axisymmetric}, an integral relation is obtained as
\begin{align}
	\hat{\F}\cdot\U_0= -\int_S \u_s\cdot(\n\cdot\hbsigma)\;dS,
\end{align}
which relates the force on the translating particle in the auxiliary problem $\hat{\F}$ with the unknown propulsion velocity $\U_0$ via a surface integral involving the surface velocity $\u_s$. By using the known solution to the auxiliary problem \citep{happel2012low} and simplifying the surface integral in the prolate spheroidal coordinates, the propulsion velocity $\U_0 = U_0 \mathbf{e}_z$ is obtained in terms of the integral \citep{popescu2010phoretic, poehnl2020axisymmetric}
\begin{align}\label{eq:U0}
	U_0 =- \frac{\tau_0}{2}\int_{-1}^1 u_s\frac{\sqrt{1-\zeta^2}}{\sqrt{\tau_0^2-\zeta^2}}\;d\zeta \cdot
\end{align}
Upon substituting the slip velocity given by \eqref{eq:ussol}, one obtain the final expression of the propulsion speed for $N$ phoretic modes as  \citep{poehnl2020axisymmetric}
\begin{equation}
    U_0 = \frac{\tau_0^2}{2} \sum_{n=1, \ \text{odd $n$}}^{N} B_n \int_{-1}^1 \frac{P_1^1(\zeta)P_n^1(\zeta)}{\tau_0^2 - \zeta^2} \ d\zeta \cdot \label{eqn:finalU0}
\end{equation}

To determine the velocity field, we consider a streamfunction $\psi_0$ for the axisymmetric flow in the co-moving frame
\begin{align}
\u_0-\U_0 = \frac{1}{h_\zeta h_\phi}\frac{\partial\psi_0}{\partial\zeta} \e_\tau- \frac{1}{h_\tau h_\phi}\frac{\partial\psi_0}{\partial\tau} \e_\zeta,
\end{align}
where the far-field corresponds to a uniform flow given by $-\U_0$. A general solution of the streamfunction can be expanded in terms of products of the Gegenbauer functions in the prolate spheroidal coordinates. For a bounded solution satisfying the far-field and force-free condition, the streamfunction takes the form \citep{poehnl2020axisymmetric}
 \begin{align}\label{eq:psi0}
 \psi_0(\tau,\zeta)= \sum_{n=2}^\infty g_n(\tau)G_n(\zeta),	
\end{align}
with $g_2(\tau) = C_4 H_4(\tau)+D_2H_2(\tau)-2c^2U_0G_2(\tau)$, $g_3(\tau) = C_3+C_5H_5(\tau)+D_3H_3(\tau)$, and $g_n(\tau) = C_{n+2}H_{n+2}(\tau)+C_nH_{n-2}(\tau)+D_nH_n(\tau)$ for $n\geq 4$, 
where $G_n$ and $H_n$ are the Gegenbauer polynomials of the first and the second kinds, respectively. The coefficients $C_n$ and $D_n$ are determined by the tangential slip velocity and the zero normal velocity on the particle surface. If we consider a slip velocity (\ref{eq:ussol}) expansion of only $N$ modes and apply the boundary conditions to the stream function expansion (\ref{eq:psi0}) with only  $N$ terms ranging from $n = 2$ to $n = N+1$, the following system of equations is obtained 
\begin{align}
    g_n(\tau_0) &= 0, \quad  \text{for} \ 2\leq n \leq N+1
    \label{system of equations 1},\\
    \pdv{g_n}{\tau} \Big|_{\tau = \tau_0} &=  \tau_0 c^2 n(n-1)B_{n-1}\label{system of equations 2}, \quad \text{for} \  2\leq n \leq N+1.
\end{align}
These equations can be separated into a system with coefficients that only have even indices and a system with coefficients that only have only odd indices, as all the indices of $C_n$ and $D_n$ in $g_n$ are always the same parity \citep{poehnl2020axisymmetric}. The odd system of equations always has one more unknown than equation; a solve-able system of equations is obtained by setting
\begin{equation}
    C_{N+3-m} = 0, \ \text{where} \ m= 
    \begin{cases} 
        1 \ \text{, when $N$ is odd} \\
        0 \ \text{, when $N$ is even} 
    \end{cases},
    \label{system of equations 3}
\end{equation}
for $N \ge 2$. Upon obtaining the phoretic modes $B_n$ using (\ref{eq:concentration coefficients}) and (\ref{eq:modes}) and using the result given by \eqref{eqn:finalU0}, the system of equations (\ref{system of equations 1})--(\ref{system of equations 2}) is solved for the coefficients $C_n$ and $D_n$ for the zeroth-order velocity field. Interested readers are referred to previous works for further details \citep{popescu2010phoretic, poehnl2020axisymmetric}.

\section{Validation of numerical simulations}\label{sec:appenB_Numerical}

\begin{figure}
	\centering
	\includegraphics[scale=0.45]{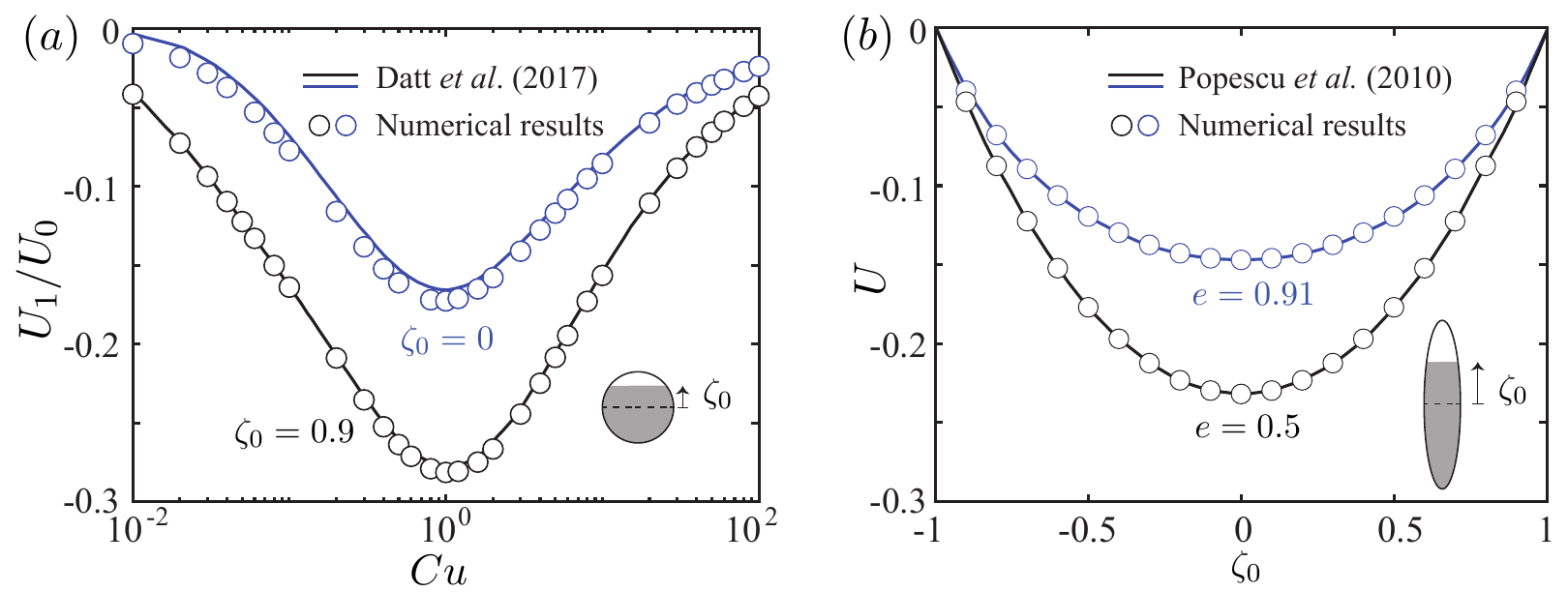}
	\caption{\footnotesize Validation of numerical approach against previously reported results. $(a)$ Comparison of the numerical results (symbols) on the swimming velocity of a spherical Janus particle as a function of $Cu$ for different active surface coverage $\zeta_0$ with the asymptotic solution (lines) of the scaled first-order swimming velocity ($U_1/U_0$) obtained by \citet{datt2017active} in a weakly shear-thinning fluid ($\beta =0.99$). $(b)$ Comparison of the numerical results (symbols) on the swimming velocity $U$ of a spheroidal Janus particle as a function of $\zeta_0$ for different eccentricities $e$ with the solution (lines) obtained by \citet{popescu2010phoretic} in a Newtonian fluid. Note that the Janus particles simulated here are coated on the bottom to maintain consistency, and correspondingly, we set $A=-1$ and $M=1$ in both $(a, b)$.}
	\label{fig:Verification}
\end{figure}

In this appendix, we include results on the validation of our numerical approach against previously reported findings. First, we follow the numerical implementation described in Section \ref{sec:Numerical} to simulate the self-propulsion of a spherical Janus particle in a shear-thinning fluid and compare the numerical results with the asymptotic solution obtained by \citet{datt2017active} in the weakly nonlinear limit ($\beta=0.99$). As shown in figure \ref{fig:Verification}$(a)$, the results display satisfactory agreements for a wide range of $\Can$ for different active surface coverage, $\zeta_0$. Second, we assess the capability of our numerical approach in handling non-spherical geometries by simulating the self-propulsion of spheroidal Janus particles with different eccentricities in a Newtonian fluid and compare the numerical results with the solution obtained by \citet{popescu2010phoretic}. As shown in figure \ref{fig:Verification}$(b)$, the results again agree satisfactorily, validating our numerical implementation.

%===========================================================

% \bibliographystyle{jfm}
% \bibliography{Ref_Janus}

\begin{thebibliography}{78}
\expandafter\ifx\csname natexlab\endcsname\relax\def\natexlab#1{#1}\fi
\def\au#1{#1} \def\ed#1{#1} \def\yr#1{#1}\def\at#1{#1}\def\jt#1{\textit{#1}}
  \def\bt#1{#1}\def\bvol#1{\textbf{#1}} \def\vol#1{#1} \def\pg#1{#1}
  \def\publ#1{#1}\def\arxiv#1{#1}\def\org#1{#1}\def\st#1{\textit{#1}}

\bibitem[Anderson(1989)]{anderson1989colloid}
{\sc \au{Anderson, John~L}} \yr{1989}  \at{Colloid transport by interfacial
  forces}.  \jt{Annu. Rev. Fluid Mech.}  \bvol{21}~(1),  \pg{61--99}.

\bibitem[Asmolov {\em et~al.\/}(2022)Asmolov, Nizkaya \&
  Vinogradova]{asmolov2022self}
{\sc \au{Asmolov, Evgeny~S}, \au{Nizkaya, Tatiana~V} \& \au{Vinogradova,
  Olga~I}} \yr{2022}  \at{Self-diffusiophoresis of janus particles that release
  ions}.  \jt{Phys. Fluids}  \bvol{34}~(3).

\bibitem[Baskurt \& Meiselman(2003)]{baskurt2003blood}
{\sc \au{Baskurt, Oguz~K} \& \au{Meiselman, Herbert~J}} \yr{2003} Blood
  rheology and hemodynamics.  \bt{In {\em Seminars in thrombosis and
  hemostasis\/}}, ,  \vol{vol.~29},  \pg{pp. 435--450}. Thieme Medical
  Publishers, Inc.

\bibitem[Bechinger {\em et~al.\/}(2016)Bechinger, Di~Leonardo, L{\"o}wen,
  Reichhardt, Volpe \& Volpe]{bechinger2016active}
{\sc \au{Bechinger, Clemens}, \au{Di~Leonardo, Roberto}, \au{L{\"o}wen,
  Hartmut}, \au{Reichhardt, Charles}, \au{Volpe, Giorgio} \& \au{Volpe,
  Giovanni}} \yr{2016}  \at{Active particles in complex and crowded
  environments}.  \jt{Rev. Mod. Phys.}  \bvol{88}~(4),  \pg{045006}.

\bibitem[Bird {\em et~al.\/}(1987)Bird, Armstrong \&
  Hassager]{bird1987dynamics}
{\sc \au{Bird, Robert~Byron}, \au{Armstrong, Robert~Calvin} \& \au{Hassager,
  Ole}} \yr{1987}  \at{Dynamics of polymeric liquids. vol. 1: Fluid mechanics}
  .

\bibitem[Blake(1971)]{blake1971spherical}
{\sc \au{Blake, John~R}} \yr{1971}  \at{A spherical envelope approach to
  ciliary propulsion}.  \jt{J. Fluid Mech.}  \bvol{46}~(1),  \pg{199--208}.

\bibitem[Brown \& Poon(2014)]{Brown2014}
{\sc \au{Brown, Aidan} \& \au{Poon, Wilson}} \yr{2014}  \at{Ionic effects in
  self-propelled pt-coated janus swimmers}.  \jt{Soft Matter}  \bvol{10},
  \pg{4016--4027}.

\bibitem[Brown {\em et~al.\/}(2017)Brown, Poon, Holm \&
  De~Graaf]{brown2017ionic}
{\sc \au{Brown, Aidan~T}, \au{Poon, Wilson~CK}, \au{Holm, Christian} \&
  \au{De~Graaf, Joost}} \yr{2017}  \at{Ionic screening and dissociation are
  crucial for understanding chemical self-propulsion in polar solvents}.
  \jt{Soft Matter}  \bvol{13}~(6),  \pg{1200--1222}.

\bibitem[Buttinoni {\em et~al.\/}(2012)Buttinoni, Volpe, K{\"u}mmel, Volpe \&
  Bechinger]{buttinoni2012active}
{\sc \au{Buttinoni, Ivo}, \au{Volpe, Giovanni}, \au{K{\"u}mmel, Felix},
  \au{Volpe, Giorgio} \& \au{Bechinger, Clemens}} \yr{2012}  \at{Active
  brownian motion tunable by light}.  \jt{J. Phys.: Condens. Matter}
  \bvol{24}~(28),  \pg{284129}.

\bibitem[Champion {\em et~al.\/}(2007)Champion, Katare \&
  Mitragotri]{champion2007making}
{\sc \au{Champion, Julie~A}, \au{Katare, Yogesh~K} \& \au{Mitragotri, Samir}}
  \yr{2007}  \at{Making polymeric micro-and nanoparticles of complex shapes}.
  \jt{Proc. Natl. Acad. Sci.}  \bvol{104}~(29),  \pg{11901--11904}.

\bibitem[Champion \& Mitragotri(2006)]{champion2006role}
{\sc \au{Champion, Julie~A} \& \au{Mitragotri, Samir}} \yr{2006}  \at{Role of
  target geometry in phagocytosis}.  \jt{Proc. Natl. Acad. Sci.}
  \bvol{103}~(13),  \pg{4930--4934}.

\bibitem[Choudhary {\em et~al.\/}(2020)Choudhary, Renganathan \&
  Pushpavanam]{choudhary2020non}
{\sc \au{Choudhary, Akash}, \au{Renganathan, Thiruvengadam} \& \au{Pushpavanam,
  S}} \yr{2020}  \at{Non-newtonian effects on the slip and mobility of a
  self-propelling active particle}.  \jt{J. Fluid Mech.}  \bvol{899}.

\bibitem[Daddi-Moussa-Ider {\em et~al.\/}(2021)Daddi-Moussa-Ider, Nasouri,
  Vilfan \& Golestanian]{daddi2021optimal}
{\sc \au{Daddi-Moussa-Ider, Abdallah}, \au{Nasouri, Babak}, \au{Vilfan, Andrej}
  \& \au{Golestanian, Ramin}} \yr{2021}  \at{Optimal swimmers can be pullers,
  pushers or neutral depending on the shape}.  \jt{Journal of Fluid Mechanics}
  \bvol{922},  \pg{R5}.

\bibitem[Datt {\em et~al.\/}(2017)Datt, Natale, Hatzikiriakos \&
  Elfring]{datt2017active}
{\sc \au{Datt, Charu}, \au{Natale, Giovanniantonio}, \au{Hatzikiriakos,
  Savvas~G} \& \au{Elfring, Gwynn~J}} \yr{2017}  \at{An active particle in a
  complex fluid}.  \jt{J. Fluid Mech.}  \bvol{823},  \pg{675--688}.

\bibitem[Datt {\em et~al.\/}(2015)Datt, Zhu, Elfring \& Pak]{datt2015squirming}
{\sc \au{Datt, Charu}, \au{Zhu, Lailai}, \au{Elfring, Gwynn~J} \& \au{Pak,
  On~Shun}} \yr{2015}  \at{Squirming through shear-thinning fluids}.  \jt{J.
  Fluid Mech.}  \bvol{784}.

\bibitem[De~Corato {\em et~al.\/}(2020)De~Corato, Arqu{\'e}, Pati{\~n}o,
  Arroyo, S{\'a}nchez \& Pagonabarraga]{de2020self}
{\sc \au{De~Corato, Marco}, \au{Arqu{\'e}, Xavier}, \au{Pati{\~n}o, Tania},
  \au{Arroyo, Marino}, \au{S{\'a}nchez, Samuel} \& \au{Pagonabarraga, Ignacio}}
  \yr{2020}  \at{Self-propulsion of active colloids via ion release: Theory and
  experiments}.  \jt{Phys. Rev. Lett.}  \bvol{124}~(10),  \pg{108001}.

\bibitem[De~Corato {\em et~al.\/}(2015)De~Corato, Greco \&
  Maffettone]{de2015locomotion}
{\sc \au{De~Corato, M}, \au{Greco, F} \& \au{Maffettone, PL}} \yr{2015}
  \at{Locomotion of a microorganism in weakly viscoelastic liquids}.  \jt{Phys.
  Rev. E}  \bvol{92}~(5),  \pg{053008}.

\bibitem[Demir {\em et~al.\/}(2020)Demir, Lordi, Ding \&
  Pak]{demir2020nonlocal}
{\sc \au{Demir, Ebru}, \au{Lordi, Noah}, \au{Ding, Yang} \& \au{Pak, On~Shun}}
  \yr{2020}  \at{Nonlocal shear-thinning effects substantially enhance helical
  propulsion}.  \jt{Phys. Rev. Fluids}  \bvol{5}~(11),  \pg{111301}.

\bibitem[Ebbens {\em et~al.\/}(2014)Ebbens, Gregory, Dunderdale, Howse,
  Ibrahim, Liverpool \& Golestanian]{Ebbens_2014}
{\sc \au{Ebbens, S.}, \au{Gregory, D.~A.}, \au{Dunderdale, G.}, \au{Howse,
  J.~R.}, \au{Ibrahim, Y.}, \au{Liverpool, T.~B.} \& \au{Golestanian, R.}}
  \yr{2014}  \at{Electrokinetic effects in catalytic platinum-insulator janus
  swimmers}.  \jt{Europhys. Lett.}  \bvol{106}~(5),  \pg{58003}.

\bibitem[Elfring \& Lauga(2015)]{elfring2015theory}
{\sc \au{Elfring, Gwynn~J} \& \au{Lauga, Eric}} \yr{2015}  \at{Theory of
  locomotion through complex fluids}.  \jt{Complex Fluids Biol. Syst.: Exp.,
  Theory, Comput.}  \pg{pp. 283--317}.

\bibitem[Eloul {\em et~al.\/}(2020)Eloul, Poon, Farago \& Frenkel]{Eloul2020}
{\sc \au{Eloul, Shaltiel}, \au{Poon, Wilson C.~K.}, \au{Farago, Oded} \&
  \au{Frenkel, Daan}} \yr{2020}  \at{Reactive momentum transfer contributes to
  the self-propulsion of janus particles}.  \jt{Phys. Rev. Lett.}  \bvol{124},
  \pg{188001}.

\bibitem[Fauci \& Dillon(2006)]{Fauci06}
{\sc \au{Fauci, L.~J.} \& \au{Dillon, R.}} \yr{2006}  \at{Biofluidmechanics of
  reproduction}.  \jt{Annu. Rev. Fluid Mech.}  \bvol{38}~(1),  \pg{371--394}.

\bibitem[Gagnon {\em et~al.\/}(2014)Gagnon, Keim \&
  Arratia]{gagnon2014undulatory}
{\sc \au{Gagnon, David~A}, \au{Keim, Nathan~C} \& \au{Arratia, Paulo~E}}
  \yr{2014}  \at{Undulatory swimming in shear-thinning fluids: experiments with
  caenorhabditis elegans}.  \jt{J. Fluid Mech.}  \bvol{758},  \pg{R3}.

\bibitem[Gao \& Wang(2014)]{gao2014synthetic}
{\sc \au{Gao, Wei} \& \au{Wang, Joseph}} \yr{2014}  \at{Synthetic
  micro/nanomotors in drug delivery}.  \jt{Nanoscale}  \bvol{6}~(18),
  \pg{10486--10494}.

\bibitem[Ghosh \& Fischer(2009)]{Ghosh2009}
{\sc \au{Ghosh, A.} \& \au{Fischer, P.}} \yr{2009}  \at{Controlled propulsion
  of artificial magnetic nanostructured propellers}.  \jt{Nano Lett.}
  \bvol{9}~(6),  \pg{2243--2245}.

\bibitem[Glotzer \& Solomon(2007)]{glotzer2007anisotropy}
{\sc \au{Glotzer, Sharon~C} \& \au{Solomon, Michael~J}} \yr{2007}
  \at{Anisotropy of building blocks and their assembly into complex
  structures}.  \jt{Nat. Mater.}  \bvol{6}~(8),  \pg{557--562}.

\bibitem[van Gogh {\em et~al.\/}(2022)van Gogh, Demir, Palaniappan \&
  Pak]{van2022effect}
{\sc \au{van Gogh, Brandon}, \au{Demir, Ebru}, \au{Palaniappan, D} \& \au{Pak,
  On~Shun}} \yr{2022}  \at{The effect of particle geometry on squirming through
  a shear-thinning fluid}.  \jt{J. Fluid Mech.}  \bvol{938}.

\bibitem[Golestanian {\em et~al.\/}(2007)Golestanian, Liverpool \&
  Ajdari]{golestanian2007designing}
{\sc \au{Golestanian, Ramin}, \au{Liverpool, TB} \& \au{Ajdari, A}} \yr{2007}
  \at{Designing phoretic micro-and nano-swimmers}.  \jt{New J. Phys.}
  \bvol{9}~(5),  \pg{126}.

\bibitem[Golestanian {\em et~al.\/}(2005)Golestanian, Liverpool \&
  Ajdari]{golestanian2005propulsion}
{\sc \au{Golestanian, Ramin}, \au{Liverpool, Tanniemola~B} \& \au{Ajdari,
  Armand}} \yr{2005}  \at{Propulsion of a molecular machine by asymmetric
  distribution of reaction products}.  \jt{Phys. Rev. Lett.}  \bvol{94}~(22),
  \pg{220801}.

\bibitem[G{\'o}mez {\em et~al.\/}(2017)G{\'o}mez, God{\'\i}nez, Lauga \&
  Zenit]{gomez2017helical}
{\sc \au{G{\'o}mez, Sa{\'u}l}, \au{God{\'\i}nez, Francisco~A}, \au{Lauga, Eric}
  \& \au{Zenit, Roberto}} \yr{2017}  \at{Helical propulsion in shear-thinning
  fluids}.  \jt{J. Fluid Mech.}  \bvol{812}.

\bibitem[Guo {\em et~al.\/}(2021)Guo, Zhu, Liu, Bonnet \&
  Veerapaneni]{guo2021optimal}
{\sc \au{Guo, Hanliang}, \au{Zhu, Hai}, \au{Liu, Ruowen}, \au{Bonnet, Marc} \&
  \au{Veerapaneni, Shravan}} \yr{2021}  \at{Optimal ciliary locomotion of
  axisymmetric microswimmers}.  \jt{Journal of Fluid Mechanics}  \bvol{927},
  \pg{A22}.

\bibitem[Happel \& Brenner(2012)]{happel2012low}
{\sc \au{Happel, John} \& \au{Brenner, Howard}} \yr{2012} {\em Low Reynolds
  number hydrodynamics: with special applications to particulate media\/}, ,
  \vol{vol.~1}.  \publ{Springer Science \& Business Media}.

\bibitem[Howse {\em et~al.\/}(2007)Howse, Jones, Ryan, Gough, Vafabakhsh \&
  Golestanian]{Howse2007}
{\sc \au{Howse, Jonathan~R.}, \au{Jones, Richard A.~L.}, \au{Ryan, Anthony~J.},
  \au{Gough, Tim}, \au{Vafabakhsh, Reza} \& \au{Golestanian, Ramin}} \yr{2007}
  \at{Self-motile colloidal particles: From directed propulsion to random
  walk}.  \jt{Phys. Rev. Lett.}  \bvol{99},  \pg{048102}.

\bibitem[Hwang {\em et~al.\/}(1969)Hwang, Litt \&
  Forsman]{hwang1969rheological}
{\sc \au{Hwang, SH}, \au{Litt, M} \& \au{Forsman, WC}} \yr{1969}
  \at{Rheological properties of mucus}.  \jt{Rheol. Acta.}  \bvol{8}~(4),
  \pg{438--448}.

\bibitem[Ishimoto \& Gaffney(2013)]{ishimoto2013squirmer}
{\sc \au{Ishimoto, Kenta} \& \au{Gaffney, Eamonn~A}} \yr{2013}  \at{Squirmer
  dynamics near a boundary}.  \jt{Phys. Rev. E}  \bvol{88}~(6),  \pg{062702}.

\bibitem[Jülicher \& Prost(2009)]{Julicher2009}
{\sc \au{Jülicher, F.} \& \au{Prost, J.}} \yr{2009}  \at{Generic theory of
  colloidal transport}.  \jt{Eur. Phys. J. E}  \bvol{29},  \pg{27--36}.

\bibitem[Kagan {\em et~al.\/}(2009)Kagan, Calvo-Marzal, Balasubramanian,
  Sattayasamitsathit, Manesh, Flechsig \& Wang]{kagan2009chemical}
{\sc \au{Kagan, Daniel}, \au{Calvo-Marzal, Percy}, \au{Balasubramanian,
  Shankar}, \au{Sattayasamitsathit, Sirilak}, \au{Manesh, Kalayil~Manian},
  \au{Flechsig, Gerd-Uwe} \& \au{Wang, Joseph}} \yr{2009}  \at{Chemical sensing
  based on catalytic nanomotors: motion-based detection of trace silver}.
  \jt{J. Am. Chem. Soc.}  \bvol{131}~(34),  \pg{12082--12083}.

\bibitem[Katsamba {\em et~al.\/}(2022)Katsamba, Butler, Koens \&
  Montenegro-Johnson]{katsamba2022chemically}
{\sc \au{Katsamba, Panayiota}, \au{Butler, Matthew~D}, \au{Koens, Lyndon} \&
  \au{Montenegro-Johnson, Thomas~D}} \yr{2022}  \at{Chemically active
  filaments: analysis and extensions of slender phoretic theory}.  \jt{Soft
  Matter}  \bvol{18}~(37),  \pg{7051--7063}.

\bibitem[Keller \& Wu(1977)]{keller1977porous}
{\sc \au{Keller, Stuart~R} \& \au{Wu, Theodore~Y}} \yr{1977}  \at{A porous
  prolate-spheroidal model for ciliated micro-organisms}.  \jt{J. Fluid Mech.}
  \bvol{80}~(2),  \pg{259--278}.

\bibitem[Lauga(2014)]{lauga2014locomotion}
{\sc \au{Lauga, Eric}} \yr{2014}  \at{Locomotion in complex fluids: integral
  theorems}.  \jt{Phys. Fluids.}  \bvol{26}~(8),  \pg{081902}.

\bibitem[Lauga(2016)]{Lauga2016_Annu}
{\sc \au{Lauga, E.}} \yr{2016}  \at{Bacterial hydrodynamics}.  \jt{Annu. Rev.
  Fluid Mech.}  \bvol{48},  \pg{105--130}.

\bibitem[Lauga \& Michelin(2016)]{Lauga2016}
{\sc \au{Lauga, Eric} \& \au{Michelin, S\'ebastien}} \yr{2016}  \at{Stresslets
  induced by active swimmers}.  \jt{Phys. Rev. Lett.}  \bvol{117},
  \pg{148001}.

\bibitem[Lauga \& Powers(2009)]{lauga2009hydrodynamics}
{\sc \au{Lauga, Eric} \& \au{Powers, Thomas~R}} \yr{2009}  \at{The
  hydrodynamics of swimming microorganisms}.  \jt{Rep. Prog. Phys.}
  \bvol{72}~(9),  \pg{096601}.

\bibitem[Leshansky {\em et~al.\/}(2007)Leshansky, Kenneth, Gat \&
  Avron]{Leshansky_2007}
{\sc \au{Leshansky, A~M}, \au{Kenneth, O}, \au{Gat, O} \& \au{Avron, J~E}}
  \yr{2007}  \at{A frictionless microswimmer}.  \jt{New J. Phys.}
  \bvol{9}~(5),  \pg{145}.

\bibitem[Li \& Ardekani(2015)]{li2015undulatory}
{\sc \au{Li, Gaojin} \& \au{Ardekani, Arezoo~M}} \yr{2015}  \at{Undulatory
  swimming in non-newtonian fluids}.  \jt{J. Fluid Mech.}  \bvol{784},
  \pg{R4}.

\bibitem[Li {\em et~al.\/}(2021)Li, Lauga \& Ardekani]{li2021microswimming}
{\sc \au{Li, Gaojin}, \au{Lauga, Eric} \& \au{Ardekani, Arezoo~M}} \yr{2021}
  \at{Microswimming in viscoelastic fluids}.  \jt{J. Non-Newton. Fluid Mech.}
  \bvol{297},  \pg{104655}.

\bibitem[Lighthill(1975)]{Lighthill1975}
{\sc \au{Lighthill, J}} \yr{1975} {\em Mathematical Biofluiddynamics\/}.
  \publ{SIAM, Philadelphia}.

\bibitem[Lighthill(1952)]{lighthill1952squirming}
{\sc \au{Lighthill, MJ}} \yr{1952}  \at{On the squirming motion of nearly
  spherical deformable bodies through liquids at very small reynolds numbers}.
  \jt{Comm. Pure Appl. Math.}  \bvol{5}~(2),  \pg{109--118}.

\bibitem[Michelin \& Lauga(2014)]{michelin2014phoretic}
{\sc \au{Michelin, S{\'e}bastien} \& \au{Lauga, Eric}} \yr{2014}  \at{Phoretic
  self-propulsion at finite p{\'e}clet numbers}.  \jt{J. Fluid Mech.}
  \bvol{747},  \pg{572--604}.

\bibitem[Montenegro-Johnson {\em et~al.\/}(2013)Montenegro-Johnson, Smith \&
  Loghin]{montenegro2013physics}
{\sc \au{Montenegro-Johnson, Thomas~D}, \au{Smith, David~J} \& \au{Loghin,
  Daniel}} \yr{2013}  \at{Physics of rheologically enhanced propulsion:
  Different strokes in generalized stokes}.  \jt{Phys. Fluids.}  \bvol{25}~(8),
   \pg{081903}.

\bibitem[Moran \& Posner(2017)]{moran2017phoretic}
{\sc \au{Moran, Jeffrey~L} \& \au{Posner, Jonathan~D}} \yr{2017}  \at{Phoretic
  self-propulsion}.  \jt{Annu. Rev. Fluid Mech.}  \bvol{49},  \pg{511--540}.

\bibitem[Natale {\em et~al.\/}(2017)Natale, Datt, Hatzikiriakos \&
  Elfring]{natale2017autophoretic}
{\sc \au{Natale, Giovanniantonio}, \au{Datt, Charu}, \au{Hatzikiriakos,
  Savvas~G} \& \au{Elfring, Gwynn~J}} \yr{2017}  \at{Autophoretic locomotion in
  weakly viscoelastic fluids at finite p{\'e}clet number}.  \jt{Phys. Fluids}
  \bvol{29}~(12).

\bibitem[Park {\em et~al.\/}(2016)Park, Kim, Shin \& Weitz]{park2016efficient}
{\sc \au{Park, Jin-Sung}, \au{Kim, Daeyeon}, \au{Shin, Jennifer~H} \&
  \au{Weitz, David~A}} \yr{2016}  \at{Efficient nematode swimming in a shear
  thinning colloidal suspension}.  \jt{Soft Matter}  \bvol{12}~(6),
  \pg{1892--1897}.

\bibitem[Pati{\~n}o {\em et~al.\/}(2018)Pati{\~n}o, Arqu{\'e}, Mestre, Palacios
  \& S{\'a}nchez]{patino2018fundamental}
{\sc \au{Pati{\~n}o, Tania}, \au{Arqu{\'e}, Xavier}, \au{Mestre, Rafael},
  \au{Palacios, Lucas} \& \au{S{\'a}nchez, Samuel}} \yr{2018}  \at{Fundamental
  aspects of enzyme-powered micro-and nanoswimmers}.  \jt{Acc. Chem. Res.}
  \bvol{51}~(11),  \pg{2662--2671}.

\bibitem[Patteson {\em et~al.\/}(2016)Patteson, Gopinath \&
  Arratia]{patteson2016active}
{\sc \au{Patteson, Alison~E}, \au{Gopinath, Arvind} \& \au{Arratia, Paulo~E}}
  \yr{2016}  \at{Active colloids in complex fluids}.  \jt{Curr. Opin. Colloid
  Interface Sci.}  \bvol{21},  \pg{86--96}.

\bibitem[Pedley(2016)]{Pedley16}
{\sc \au{Pedley, T.~J.}} \yr{2016}  \at{Spherical squirmers: Models for
  swimming micro-organisms}.  \jt{IMA J. Appl. Math.}  \bvol{81},
  \pg{488--521}.

\bibitem[Poehnl {\em et~al.\/}(2020)Poehnl, Popescu \&
  Uspal]{poehnl2020axisymmetric}
{\sc \au{Poehnl, Ruben}, \au{Popescu, Mihail~N} \& \au{Uspal, William~E}}
  \yr{2020}  \at{Axisymmetric spheroidal squirmers and self-diffusiophoretic
  particles}.  \jt{J. Phys.: Condens. Matter.}  \bvol{32}~(16),  \pg{164001}.

\bibitem[Poehnl \& Uspal(2021)]{poehnl2021phoretic}
{\sc \au{Poehnl, Ruben} \& \au{Uspal, William}} \yr{2021}  \at{Phoretic
  self-propulsion of helical active particles}.  \jt{J. Fluid Mech.}
  \bvol{927},  \pg{A46}.

\bibitem[Popescu {\em et~al.\/}(2010)Popescu, Dietrich, Tasinkevych \&
  Ralston]{popescu2010phoretic}
{\sc \au{Popescu, Mihail~Nicolae}, \au{Dietrich, S}, \au{Tasinkevych, M} \&
  \au{Ralston, J}} \yr{2010}  \at{Phoretic motion of spheroidal particles due
  to self-generated solute gradients}.  \jt{Eur. Phys. J. E}  \bvol{31}~(4),
  \pg{351--367}.

\bibitem[Purcell(1977)]{purcell1977life}
{\sc \au{Purcell, Edward~M}} \yr{1977}  \at{Life at low reynolds number}.
  \jt{Am. J. Phys.}  \bvol{45}~(1),  \pg{3--11}.

\bibitem[Qin {\em et~al.\/}(2021)Qin, Peng, Chen, Nganguia, Zhu \&
  Pak]{qin2021propulsion}
{\sc \au{Qin, Ke}, \au{Peng, Zhiwei}, \au{Chen, Ye}, \au{Nganguia, Herve},
  \au{Zhu, Lailai} \& \au{Pak, On~Shun}} \yr{2021}  \at{Propulsion of an
  elastic filament in a shear-thinning fluid}.  \jt{Soft Matter}
  \bvol{17}~(14),  \pg{3829--3839}.

\bibitem[Qu \& Breuer(2020)]{qu2020effects}
{\sc \au{Qu, Zijie} \& \au{Breuer, Kenneth~S}} \yr{2020}  \at{Effects of
  shear-thinning viscosity and viscoelastic stresses on flagellated bacteria
  motility}.  \jt{Phys. Rev. Fluids.}  \bvol{5}~(7),  \pg{073103}.

\bibitem[Saad \& Natale(2019)]{saad2019diffusiophoresis}
{\sc \au{Saad, Shabab} \& \au{Natale, Giovanniantonio}} \yr{2019}
  \at{Diffusiophoresis of active colloids in viscoelastic media}.  \jt{Soft
  Matter}  \bvol{15}~(48),  \pg{9909--9919}.

\bibitem[S{\'a}nchez {\em et~al.\/}(2015)S{\'a}nchez, Soler \&
  Katuri]{sanchez2015chemically}
{\sc \au{S{\'a}nchez, Samuel}, \au{Soler, Llu{\'\i}s} \& \au{Katuri, Jaideep}}
  \yr{2015}  \at{Chemically powered micro-and nanomotors}.  \jt{Angew. Chem.
  Int. Ed.}  \bvol{54}~(5),  \pg{1414--1444}.

\bibitem[Schwarz-Linek {\em et~al.\/}(2012)Schwarz-Linek, Valeriani, Cacciuto,
  Cates, Marenduzzo, Morozov \& Poon]{schwarz2012phase}
{\sc \au{Schwarz-Linek, J}, \au{Valeriani, C}, \au{Cacciuto, A}, \au{Cates,
  ME}, \au{Marenduzzo, D}, \au{Morozov, AN} \& \au{Poon, WCK}} \yr{2012}
  \at{Phase separation and rotor self-assembly in active particle suspensions}.
   \jt{Proc. Natl. Acad. Sci.}  \bvol{109}~(11),  \pg{4052--4057}.

\bibitem[Schweitzer \& Farmer(2003)]{schweitzer2003brownian}
{\sc \au{Schweitzer, Frank} \& \au{Farmer, J~Doyne}} \yr{2003} {\em Brownian
  agents and active particles: collective dynamics in the natural and social
  sciences\/}, ,  \vol{vol.~1}.  \publ{Springer}.

\bibitem[Shemi \& Solomon(2018)]{shemi2018self}
{\sc \au{Shemi, Onajite} \& \au{Solomon, Michael~J}} \yr{2018}
  \at{Self-propulsion and active motion of janus ellipsoids}.  \jt{J. Phys.
  Chem. B.}  \bvol{122}~(44),  \pg{10247--10255}.

\bibitem[Spagnolie \& Underhill(2023)]{spagnolie2023swimming}
{\sc \au{Spagnolie, Saverio~E} \& \au{Underhill, Patrick~T}} \yr{2023}
  \at{Swimming in complex fluids}.  \jt{Annu. Rev. Condens. Matter Phys.}
  \bvol{14},  \pg{381--415}.

\bibitem[Stone \& Samuel(1996)]{stone1996propulsion}
{\sc \au{Stone, Howard~A} \& \au{Samuel, Aravinthan~DT}} \yr{1996}
  \at{Propulsion of microorganisms by surface distortions}.  \jt{Phys. Rev.
  Lett.}  \bvol{77}~(19),  \pg{4102}.

\bibitem[Sznitman \& Arratia(2014)]{sznitman2014locomotion}
{\sc \au{Sznitman, Josu{\'e}} \& \au{Arratia, Paulo~E}} \yr{2014}
  \at{Locomotion through complex fluids: an experimental view}.  \bt{In {\em
  Complex Fluids Biol. Syst.: Exp., Theory, Comput.\/}},  \pg{pp. 245--281}.
  \publ{Springer}.

\bibitem[Theers {\em et~al.\/}(2016)Theers, Westphal, Gompper \&
  Winkler]{theers2016modeling}
{\sc \au{Theers, Mario}, \au{Westphal, Elmar}, \au{Gompper, Gerhard} \&
  \au{Winkler, Roland~G}} \yr{2016}  \at{Modeling a spheroidal microswimmer and
  cooperative swimming in a narrow slit}.  \jt{Soft Matter}  \bvol{12}~(35),
  \pg{7372--7385}.

\bibitem[V{\'e}lez-Cordero \& Lauga(2013)]{velez2013waving}
{\sc \au{V{\'e}lez-Cordero, J~Rodrigo} \& \au{Lauga, Eric}} \yr{2013}
  \at{Waving transport and propulsion in a generalized newtonian fluid}.
  \jt{J. Non-Newton. Fluid Mech.}  \bvol{199},  \pg{37--50}.

\bibitem[Wensink {\em et~al.\/}(2014)Wensink, Kantsler, Goldstein \&
  Dunkel]{wensink2014controlling}
{\sc \au{Wensink, HH}, \au{Kantsler, V}, \au{Goldstein, RE} \& \au{Dunkel, J}}
  \yr{2014}  \at{Controlling active self-assembly through broken particle-shape
  symmetry}.  \jt{Phys. Rev. E}  \bvol{89}~(1),  \pg{010302}.

\bibitem[Yariv(2019)]{yariv2019self}
{\sc \au{Yariv, Ehud}} \yr{2019}  \at{Self-diffusiophoresis of slender
  catalytic colloids}.  \jt{Langmuir}  \bvol{36}~(25),  \pg{6903--6915}.

\bibitem[Zhang {\em et~al.\/}(2009)Zhang, Abbott, Dong, Peyer, Kratochvil,
  Zhang, Bergeles \& Nelson]{Zhang2009b}
{\sc \au{Zhang, L.}, \au{Abbott, J.~J.}, \au{Dong, L.}, \au{Peyer, K.~E.},
  \au{Kratochvil, B.~E.}, \au{Zhang, H.}, \au{Bergeles, C.} \& \au{Nelson,
  B.~J.}} \yr{2009}  \at{Characterizing the swimming properties of artificial
  bacterial flagella}.  \jt{Nano Lett.}  \bvol{9}~(10),  \pg{3663--3667}.

\bibitem[Zhou {\em et~al.\/}(2018)Zhou, Zhang, Tang \&
  Wang]{zhou2018photochemically}
{\sc \au{Zhou, Chao}, \au{Zhang, HP}, \au{Tang, Jinyao} \& \au{Wang, Wei}}
  \yr{2018}  \at{Photochemically powered agcl janus micromotors as a model
  system to understand ionic self-diffusiophoresis}.  \jt{Langmuir}
  \bvol{34}~(10),  \pg{3289--3295}.

\bibitem[Zhu \& Zhu(2023)]{zhu2023self}
{\sc \au{Zhu, Guangpu} \& \au{Zhu, Lailai}} \yr{2023}  \at{Self-propulsion of
  an elliptical phoretic disk emitting solute uniformly}.  \jt{J. Fluid Mech.}
  \bvol{974},  \pg{A57}.

\bibitem[Z{\"o}ttl \& Yeomans(2019)]{zottl2019enhanced}
{\sc \au{Z{\"o}ttl, Andreas} \& \au{Yeomans, Julia~M}} \yr{2019}  \at{Enhanced
  bacterial swimming speeds in macromolecular polymer solutions}.  \jt{Nat.
  Phys.}  \bvol{15}~(6),  \pg{554--558}.

\end{thebibliography}

\end{document}